\documentclass{emulateapj}
\usepackage{amsmath}
\usepackage{times}
\usepackage{graphicx}
\usepackage{float}
\restylefloat{figure}
\usepackage[caption = false]{subfig}

\begin{document}

\newcommand{\halpha}{\ensuremath{\mbox{H}\alpha}}
\newcommand{\ith}{\ensuremath{^{\rm th}}}
\newcommand{\jk}{$J$ -- $K$}
\newcommand{\teff}{\ensuremath{T_{\mbox{\scriptsize eff}}}}
\newcommand{\logg}{\ensuremath{\log g}}
\newcommand{\persec}{\ensuremath{\mbox{s}^{-1}}}
\newcommand{\degree}{\ensuremath{^\circ}}
\newcommand{\astrosun}{\odot}
\newcommand{\msun}{\ensuremath{\mbox{M}_{\astrosun}}}
\newcommand{\vsini}{\ensuremath{v \sin i}}
\newcommand{\kms}{\ensuremath{\mbox{km s}^{-1}}}
\newcommand{\mps}{\ensuremath{\mbox{m s}^{-1}}}
\newcommand{\mas}{\ensuremath{\mbox{mas yr}^{-1}}}
\newcommand{\loden}{Lod\'{e}n}

\title{The putative old, nearby cluster Lod\'{e}n 1 does not exist \footnote{Based on observations made with the Southern African Large Telescope (SALT)}}
\author{Eunkyu Han \altaffilmark{1,2,3}}
\author{Jason L. Curtis \altaffilmark{1,2,4}}
\author{Jason T. Wright \altaffilmark{1,2}}
\slugcomment{Received by \textit{AJ} 2015 April 2; accepted 2016 April 5}

\altaffiltext{1}{Department of Astronomy and Astrophysics,
                The Pennsylvania State University,
                525 Davey Laboratory, University Park, PA 16802, USA} 
\altaffiltext{2}{Center for Exoplanets and Habitable Worlds, 
                The Pennsylvania State University,
                525 Davey Laboratory, University Park, PA 16802, USA} 
\altaffiltext{3}{Department of Astronomy, 
                Boston University, 
                725 Commonwealth Avenue, Boston, MA 02215, USA}
\altaffiltext{4}{Harvard-Smithsonian Center for Astrophysics,
                60 Garden Street, Cambridge, MA 02138, USA}
\email{eunkyuh@bu.edu}

\begin{abstract}
Astronomers have access to precious few nearby, middle-aged benchmark star clusters. Within 500 pc, there are only NGC 752 and Ruprecht 147 (R147), at 1.5 and 3 Gyr respectively. The Database for Galactic Open Clusters (WEBDA)
also lists \loden\ 1 as a 2 Gyr cluster at a distance of 360 pc.
If this is true, \loden\ 1 could become a useful benchmark cluster. 
This work details our investigation of \loden\ 1.
We assembled archival astrometry (PPMXL) and photometry (2MASS, Tycho-2, APASS), 
and acquired medium resolution spectra for radial velocity measurements with the
Robert Stobie Spectrograph (RSS) at the Southern African Large Telescope.
We observed no sign of a cluster main-sequence turnoff or red giant branch 
amongst all stars in the field brighter than $J < 11$. 
Considering the 29 stars identified by L.O. \loden\ and listed on SIMBAD as the members of \loden\ 1, we found no compelling evidence of kinematic clustering in proper motion or radial velocity.
Most of these candidates are A stars and red giants,  
and their observed properties are consistent with distant field stars 
in the direction of \loden\ 1 in the Galactic plane.
We conclude that the old nearby cluster \loden\ 1 is 
neither old, nor nearby, nor a cluster.
\end{abstract}


\section{Introduction}\label{s:intro}
There exists a loose grouping of stars located between the constellations Vela and Carina in the Southern sky\footnote{$(\alpha , \delta )$ = $10^{h}04^{m}40^{s}$, $-55^{\circ}50'00''$} that show an ``evident concentration of late-type stars...and main-sequence A-F stars,'' and whose physical association was ``not confirmed but strongly suspected" 
by \citet{loden-mem}, who dubbed these stars \loden\ 1. 
Although the reality of this open cluster has still not been proven 36 year later, 
the Database for Galactic Open Clusters 
\citep[WEBDA,\footnote{Not updated since 2013: http://www.univie.ac.at/webda/}][]{webda}
quotes the cluster's age at 2 Gyr at a distance of 360 pc.
If this is accurate, \loden\ 1 could become an important benchmark cluster for stellar astrophysics.

\begin{figure}[htb]\begin{center}
    \epsscale{1.2}
    \includegraphics[width=0.5\textwidth]{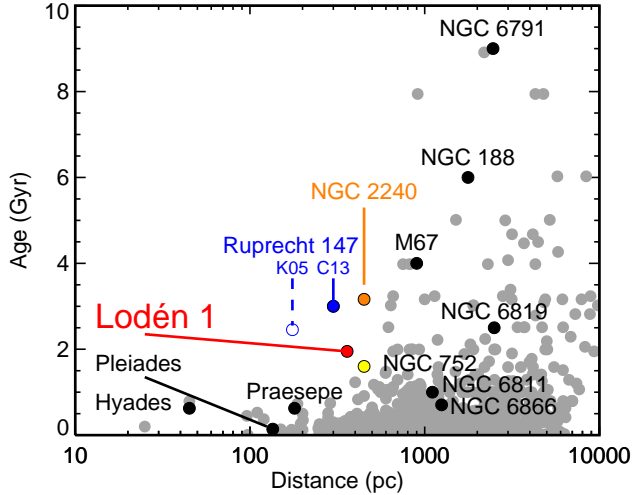}
    \caption{Ages and distances for all Galactic star clusters in WEBDA.
    At 3 Gyr and 300 pc, R147 is clearly the oldest nearby cluster 
    (blue; open circle---\citealt{khar2005}, 
    filled circle---\citealt{Curtis2013}).
    NGC 752 (1.5 Gyr, 457 pc) is marked with a yellow circle. 
    Popular benchmarks are labelled, 
    and their properties have been updated to reflect 
    the latest findings in the literature (e.g., the age of M67 is not 2.6 Gyr).
    Notice that \loden\ 1 and NGC 2240 appear to be 
    older than 1 Gyr and within 500 pc \citep{khar2005}, 
    making them promising targets for stellar astrophysics.
    \label{f:clusters}}
\end{center}\end{figure}

Astronomers have access to very few such benchmarks.
This is illustrated by Figure \ref{f:clusters}, 
which shows the dearth of nearby clusters older than 1 Gyr.
The reader will recognize a few well known nearby clusters:  
the Pleiades, Hyades, and Praesepe. These three are all within 200 pc of Earth, 
but their ages only range from 130 to 625 Myr. 
As for the prototypical old open cluster, M67 \citep{classicM67}, 
it is nearly 1 kpc away.  

While 1 kpc might seem like Earth's backyard to many astronomers, 
the calibration of empirical relationships 
for stellar ages based on rotation \citep[gyrochronology, ][]{Barnes2003} 
and magnetic activity \citep{mamajek2008}
require \textit{nearby} 
intermediate and old aged benchmarks \citep{soderblom2010}. 
Those of us that seek to study the rotation and activity of old red dwarfs 
will find M67's beyond the reach of \textit{K2} (the re-purposed \textit{Kepler} mission) 
and current ground-based optical facilities. 
M67's solar analogs also let us glimpse the long-term behavior of the Sun 
on timescales beyond even the four centuries contained in the sunspot record; 
however, any attempt to measure coronal X-rays from these analogs would 
realistically require a \textit{Chandra} ACIS integration time 
on par with, or even longer than, the \textit{Chandra} Deep Field South \citep{Brandt2001}.
These types of observational campaigns have been successfully conducted in the nearby 
NGC 752 \citep[][]{Giardino2008, Bowsher752} and R147 clusters 
\citep[e.g., \textit{K2} Campaign 7 GO 7035, ][]{SaarChandra}.

\citet{Twarog752} determined an age and distance for NGC 752 of 1.45 Gyr and 457 pc. 
As for the newly minted benchmark, Ruprecht 147 (R147), 
\citet{Curtis2013} demonstrated that it is the oldest nearby star cluster, 
with an age of 3 Gyr at a distance of only 300 pc.
This recent realization suggests that there may be other clusters 
like R147 that have been similarly overlooked.  
The utility of such clusters for stellar astrophysics demands that we find them.  
\textit{Gaia} promises to enable an 
exhaustive search for even the sparsest moving groups \citep{gaia}, 
but clusters with a statistically useful number of stars 
should be present and identifiable in existing catalogs.

R147 first caught our eye because \citet{khar2005} 
estimated its age at 2.5 Gyr and placed it at 175 pc. 
Figure \ref{f:clusters} highlights two other objects with similar
properties as estimated by their analysis:  
\loden\ 1 at 2 Gyr and 360 pc, 
and NGC 2240 at 3.2 Gyr and 450 pc.
\citet{khar2005} utilized Tycho-2 astrometry and photometry \citep{tycho2}, 
and known/suspected cluster locations, 
to automatically characterize open cluster populations and 
identify their membership.
After our investigation of \loden\ 1 was underway,
\citeauthor{khar13} repeated their cluster search and characterization program 
based on the PPMXL astrometric catalog \citep{ppmxl}
and released a new cluster database that revised \loden\ 1's age down to only 200 Myr, 
with a distance of 786 pc and $A_V = 0.22$ \citep{khar13}.   
The properties of NGC 2240 were revised to 1.58 Gyr and 1551 pc.\footnote{
\citet{revisedNewCatalog} referred to NGC 2240 as ``non-existent,''
a claim that was propagated to the \citet{dias2002} online catalog. 
Establishing the nature of these clusters matters 
and need not be postponed until the \textit{Gaia} era. 
NGC 2240 has already been incorporated into the APOGEE cluster 
survey to infer the Galactic metallicity gradient based on 
a single observed star, whose membership was inferred simply 
by proximity to the \citet{khar13} isochrone solution, 
which quite possibly modeled the Galactic field and nothing more
\citep[see OCCAM Survey, ][]{OCCAM}.} 
The large discrepancy in the \citeauthor{khar2005} results 
underscores the need for more thorough 
investigation into these purported clusters.

This work details our investigation of \loden\ 1.
We were specifically concerned with two claims.
First, \loden\ 1 is old \textit{and} nearby \citep{khar2005}, 
meaning there exists some grouping of stars in the 
\loden\ 1 field that are gravitationally bound 
together in an open cluster within 500 pc, 
and have an age $>1$ Gyr.\footnote{Again, 
these parameter limits are not arbitrary, but are motivated by 
(1) the actual lack of known clusters older than 1 Gyr within 500 pc 
and (2) the necessity to find such clusters within 500 pc so that we can 
study their red dwarf and solar analog populations in detail,
feats that are currently impossible at larger distances.} 
We also tested the separate claim that \loden\ 1, 
as defined by \citet{loden-mem} and consisting of 29 specific stars, 
actually constitutes a real cluster.

First, we checked the brightest stars in the field 
for evidence of clustering in astrometric proper motion 
and photometric archival data (\S \ref{s:blind}). 
Next, we pursued a targeted search of the cluster as defined by 
\citet{loden-mem} for youth and proximity from 
optical and NIR photometry, 
and common space motion via PPMXL proper motions  
and radial velocities we measured with spectra 
obtained from Southern African Large Telescope
(SALT) Robert Stobie Spectrograph (RSS; Section \ref{s:survey}).
We concluded that the old nearby cluster \loden\ 1 is 
neither old, nor nearby, nor a cluster (Section \ref{s:nope}).

\section{A first look at Lod\'{e}n's Field 1}
\label{s:blind}
Historically, observers identified star clusters 
by carefully scanning the skies for apparent overdensities,
which were then cataloged and classified. 
\loden 's note on his Field 1
is reminiscent of John Herschel's 1830 description of R147
as ``a very large straggling space full of loose stars''
\citep{firstr147ref}.
While dense environments like the Pleiades and M67 are unmistakably clusters, 
sparser objects like R147 and \loden\ 1 typically must be kinematically disentangled 
from the Galactic field star population. 
In fact, the visual identifications for both were challenged in the years 
following their initial discovery, 
with \citet{dias2002} describing \loden\ 1 as a
``dubious, object considered doubtful by the DSS images inspection,''
while R147 (then known as NGC 6774) 
was noted as ``possibly not a true cluster'' by \citet{burnham}.\footnote{In fact, 
\citet{Curtis2013} demonstrated that $<$25\% of the stars brighter than $V < 9$ 
in the R147 field are cluster members. Most of the bright stars that 
signaled to John Herschel and Jaroslav Ruprecht that the field contained an open cluster
actually do constitute an asterism. Fortunately, a real and remarkable cluster 
existed amongst that asterism.}

Although it sits in the Galactic plane ($l = 281$\degree, $b = 0$\degree), 
\textit{if} \loden\ 1 is real \textit{and} located within 500 pc, 
its proximity should mitigate Galactic contamination 
by making its brightest stars -- the main-sequence turnoff, 
red giant branch (RGB), and any blue straggler stars --
stand out against the field stars. 
Even at 500 pc, all of these subpopulations should appear brighter than $J < 11$.

R147 has a confirmed age and distance that is very similar to the properties 
WEBDA provides for \loden\ 1. 
With a similar range in magnitude and location in the Galactic plane 
(R147: $l = 21$\degree, $b = -13$\degree), 
R147 provides an excellent test case and demonstration of our ability to 
identify old nearby clusters hiding amongst a dense Galactic field.

We considered all stars within 30 arcmin of the centers of \loden\ 1 and R147 
by querying the PPMXL catalog 
for objects within these regions with $J < 12$,
and found 520 and 612 entries for the \loden\ 1 and R147 fields, respectively. 
We defined two empirical control samples for each field by 
selecting an annulus of equal area encompassing the primary field and 
by selecting a neighboring 30 arcmin field at fixed Galactic latitude, 
and shifted 1$^\circ$ and 1\fdg5 degrees in longitude for \loden\ 1 and R147, respectively. 
We also generated synthetic control samples using the Besan\c{c}on population 
synthesis model for the 
Milky Way \citep{galsim}.\footnote{http://model.obs-besancon.fr/}

\begin{figure*}\begin{center}
     \epsscale{0.85}
\subfloat[]{\includegraphics[width = 2.4in]{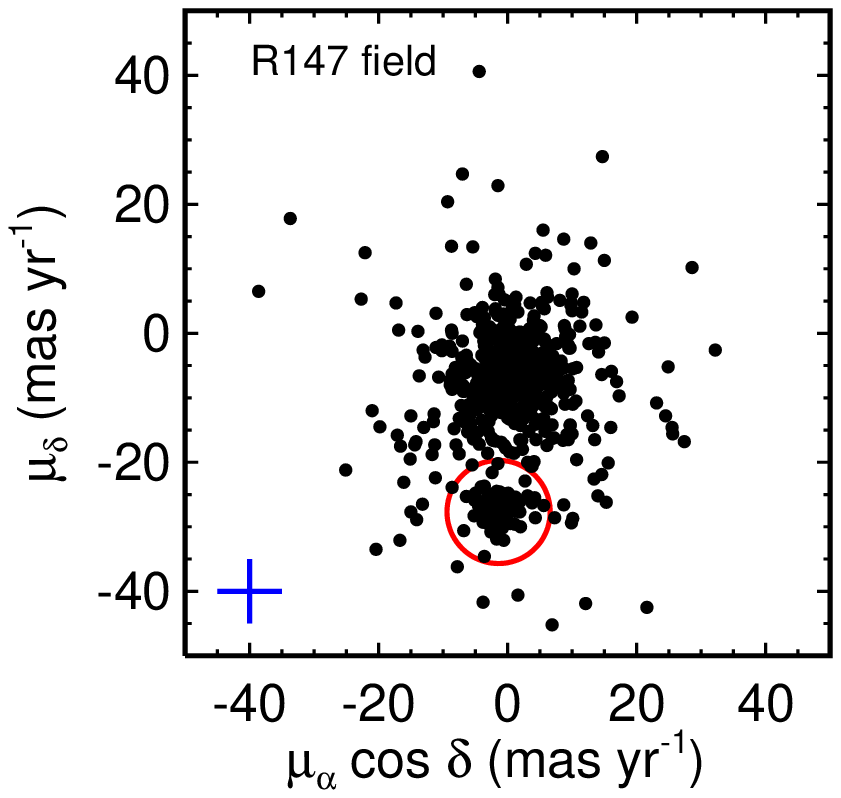}} 
\subfloat[]{\includegraphics[width = 2.4in]{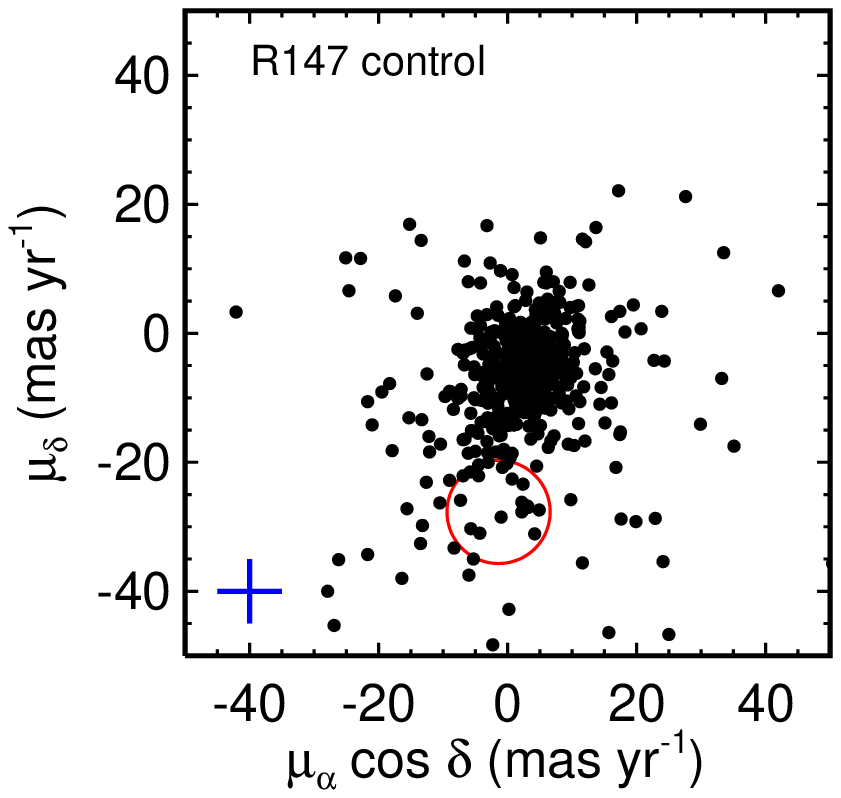}}
\subfloat[]{\includegraphics[width = 2.4in]{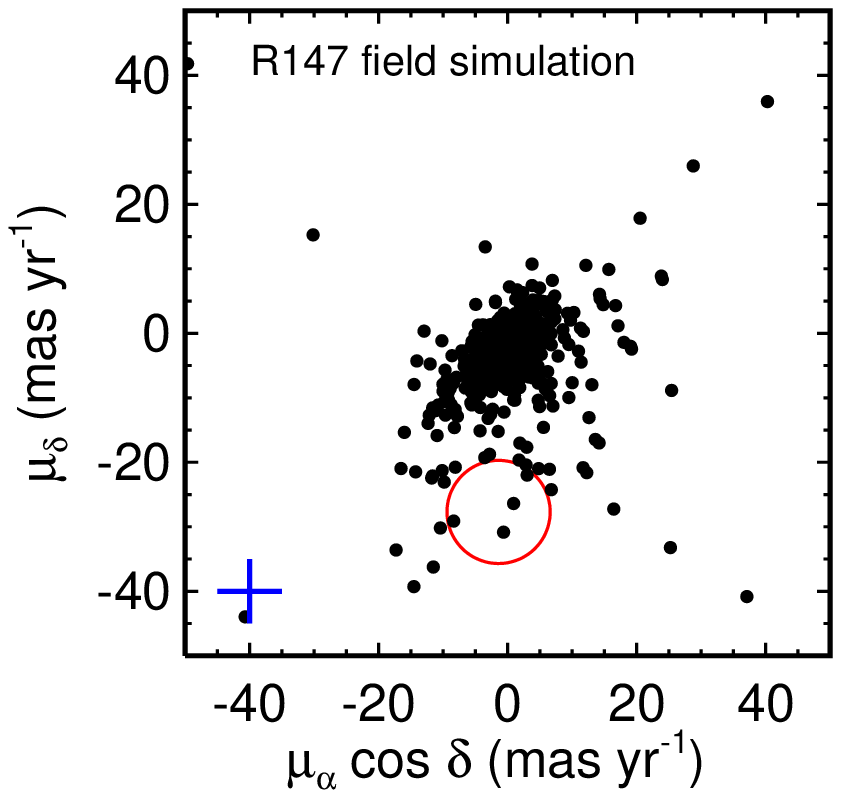}}\\
\subfloat[]{\includegraphics[width = 2.4in]{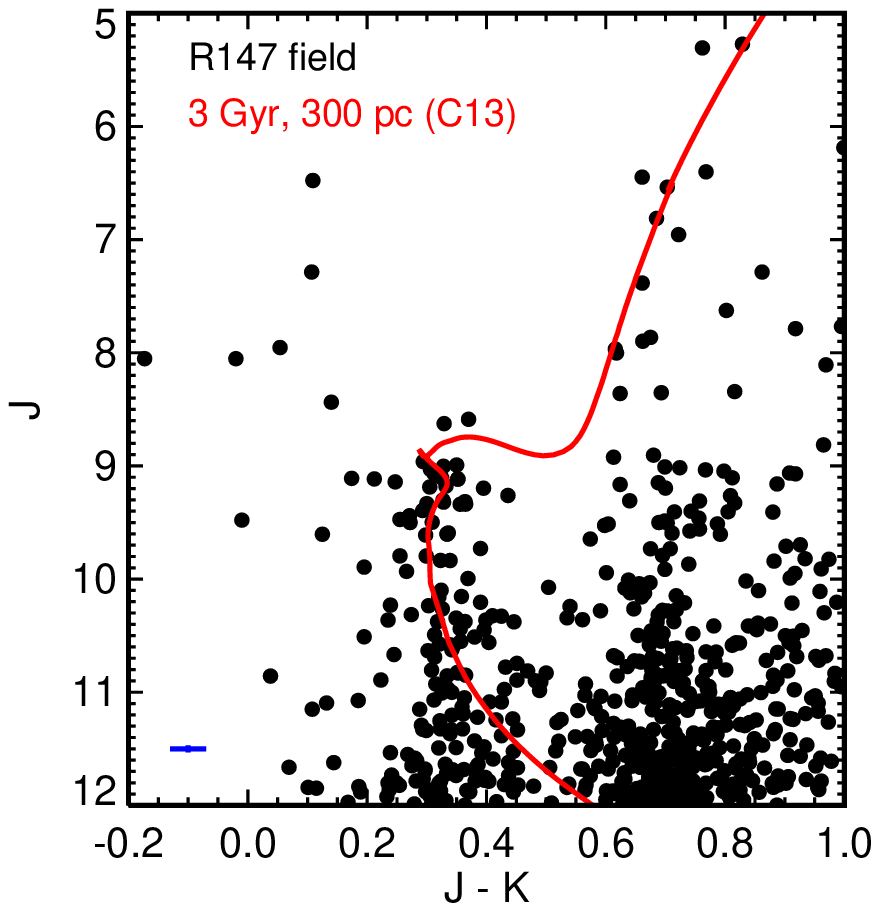}} 
\subfloat[]{\includegraphics[width = 2.4in]{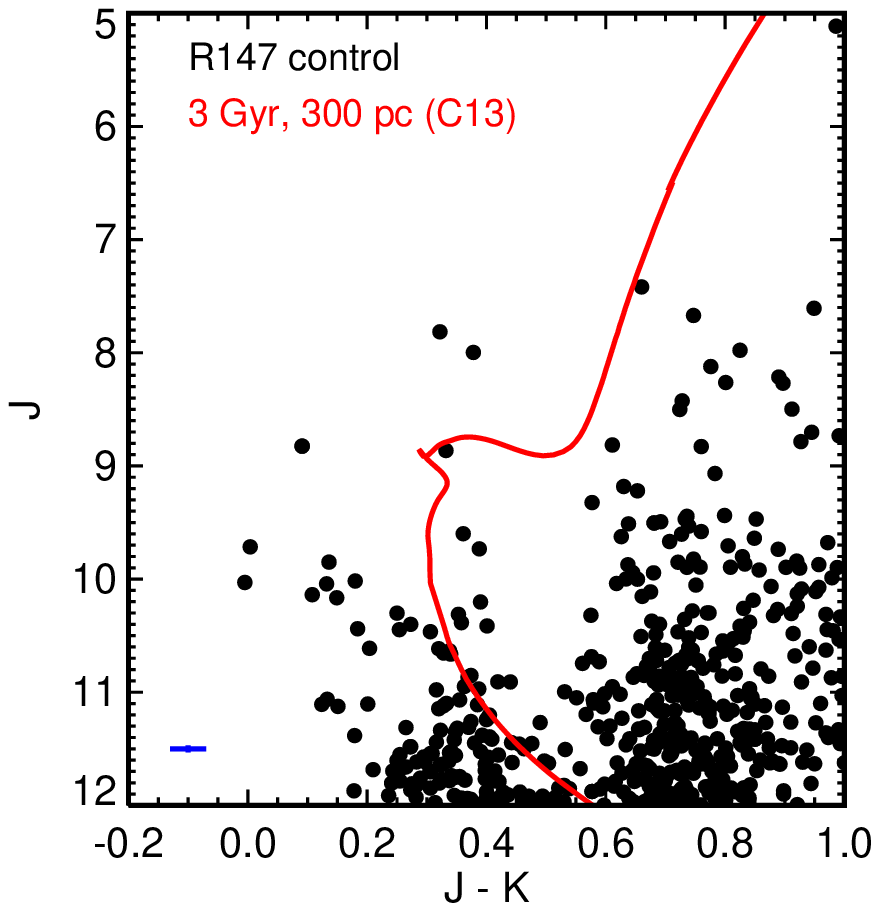}}
\subfloat[]{\includegraphics[width = 2.4in]{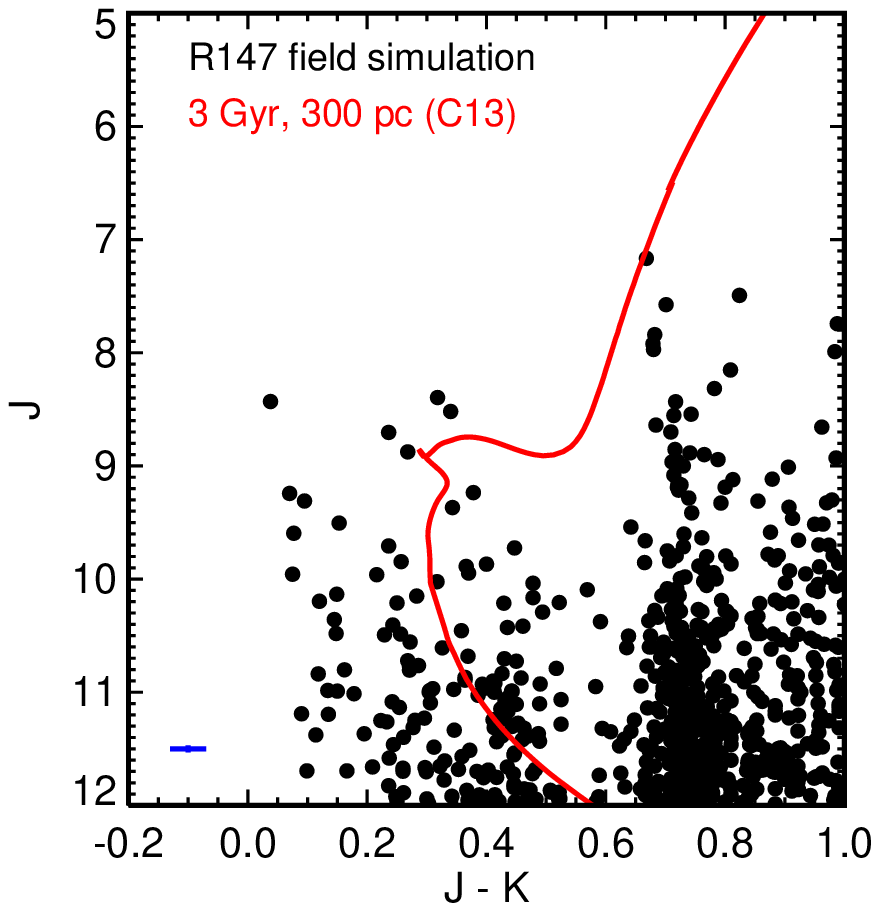}}\\
    \caption{Proper motion and color-magnitude diagrams for R147 and representative control fields.
    Left panels show all stars brighter than $J < 12$ within the innermost 30 arcmin of R147. 
    Middle panels show a neighboring field of equivalent size and magnitude cutoff
    at fixed Galactic latitude and shifted by $+$1\fdg5 in longitude. 
    Right panels show a simulation of the primary R147 field, 
    generated with the Besan\c{c}on population synthesis model for the Milky Way
    using the same location, area, and magnitude selection criteria that 
    defined the primary sample. 
    These data were queried from PPMXL.
    The red circles in the proper motion distribution panels, 
    with radii of 7 \mas , mark the R147 cluster's proper motion, 
    which cleanly separates its membership from the field at these magnitudes. 
    The blue crosses represent typical measurement uncertainties, $\sim$0.03 mag 
    for photometry, 
    and $\sim$5 \mas\ for proper motions, 
    which is why the R147 cluster members appear to fill the red circle. 
    R147 is remarkably distinct from the Galactic field star population
    in both proper motion and color--magnitude diagrams.
    }   
    \label{f:r147}
\end{center}
\end{figure*}

\begin{figure*}\begin{center}
     \epsscale{0.85}
\subfloat[]{\includegraphics[width = 2.4in]{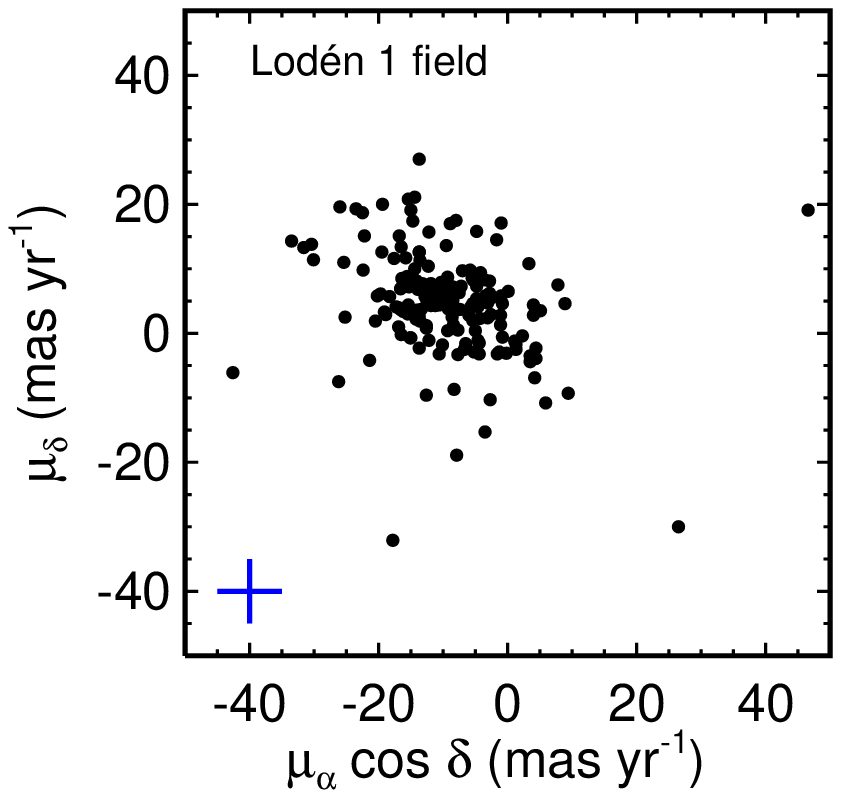}} 
\subfloat[]{\includegraphics[width = 2.4in]{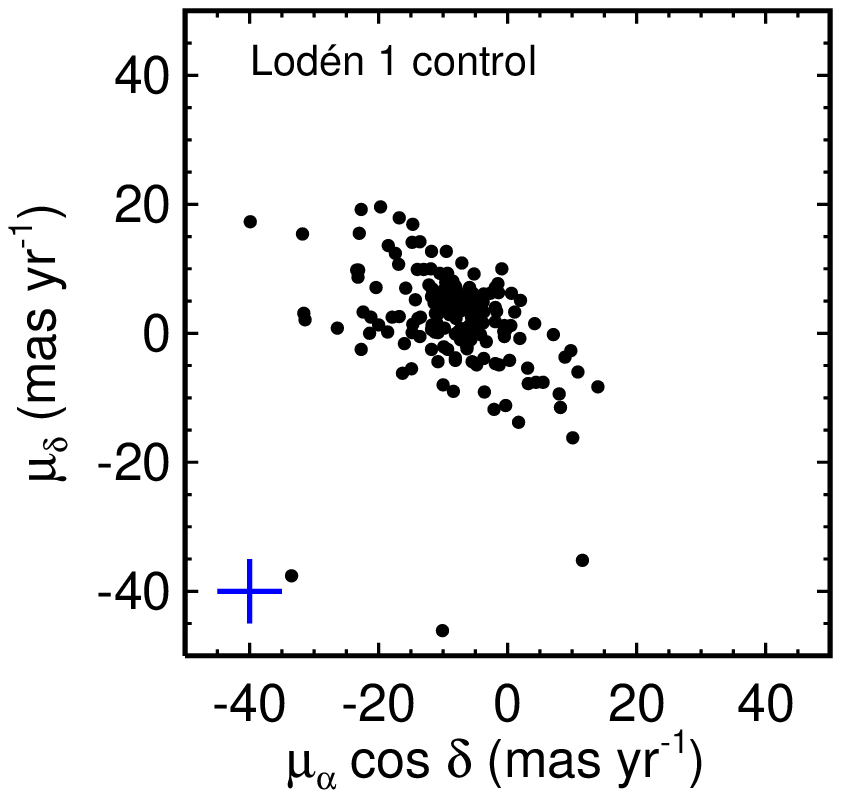}}
\subfloat[]{\includegraphics[width = 2.4in]{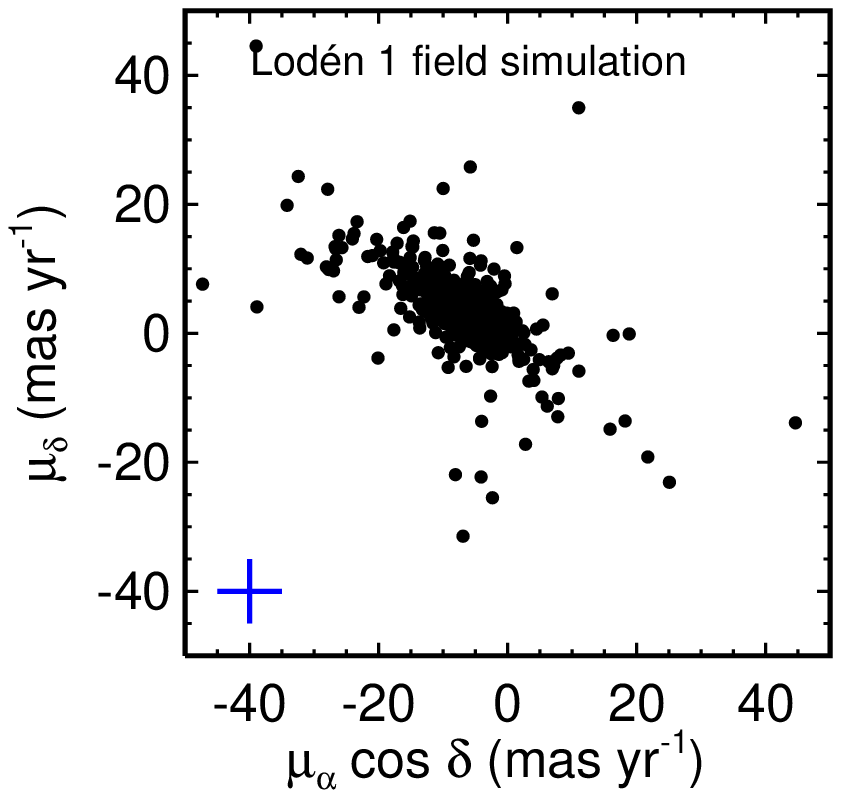}}\\
\subfloat[]{\includegraphics[width = 2.4in]{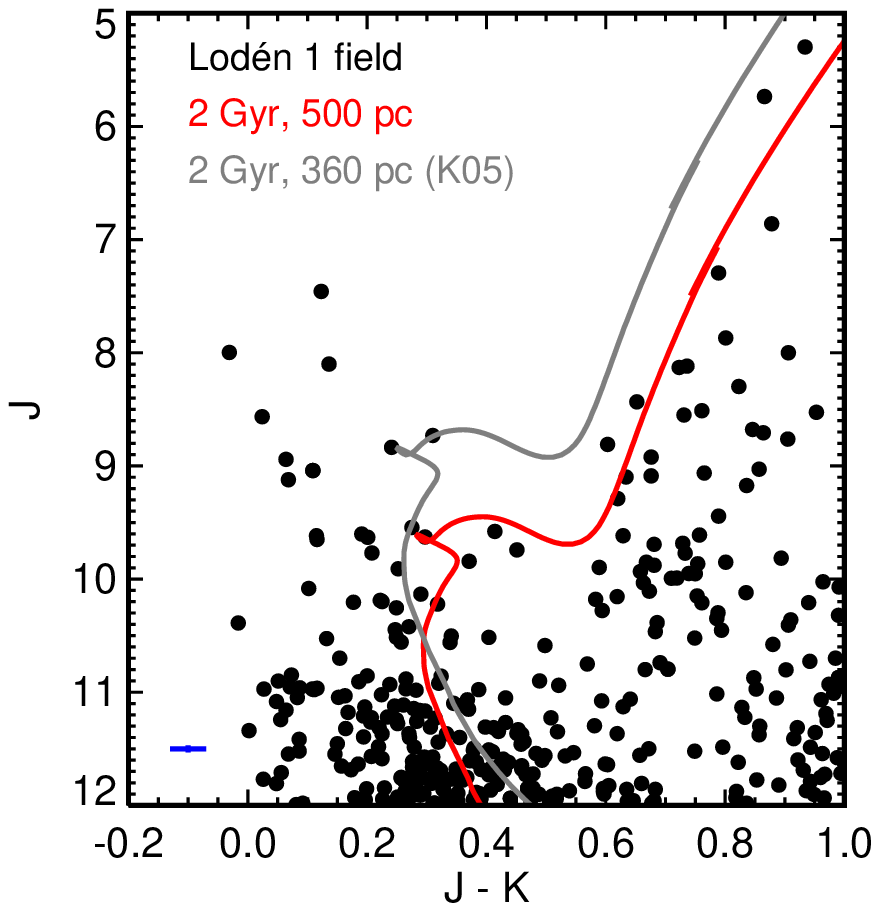}} 
\subfloat[]{\includegraphics[width = 2.4in]{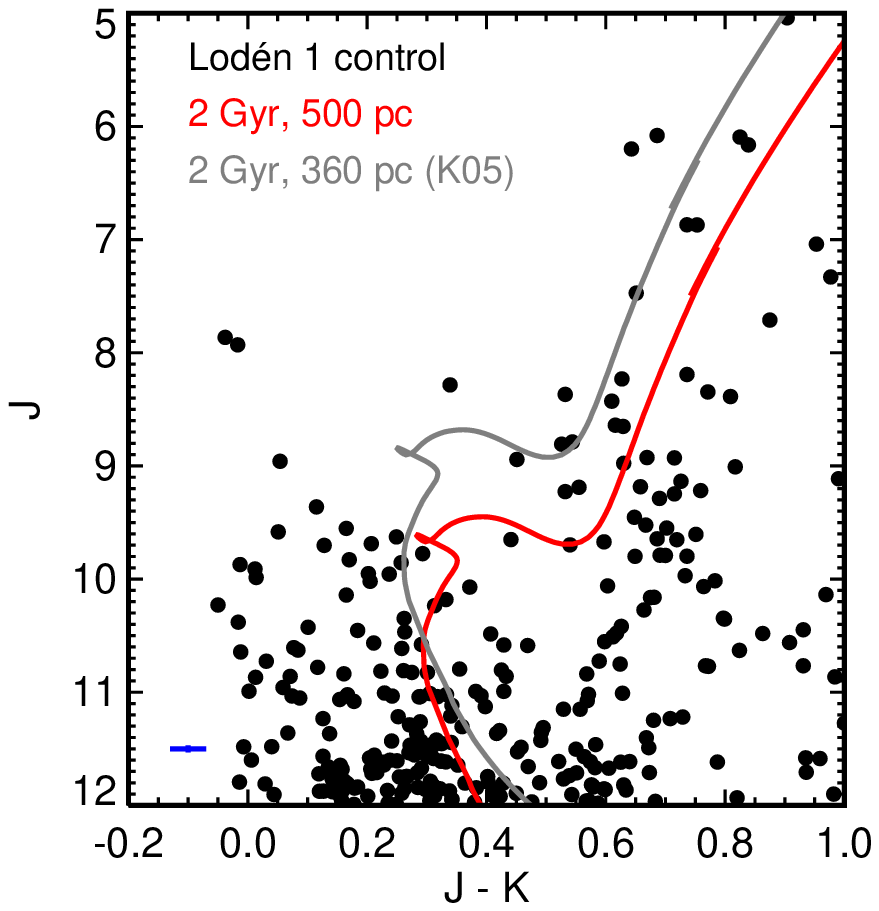}}
\subfloat[]{\includegraphics[width = 2.4in]{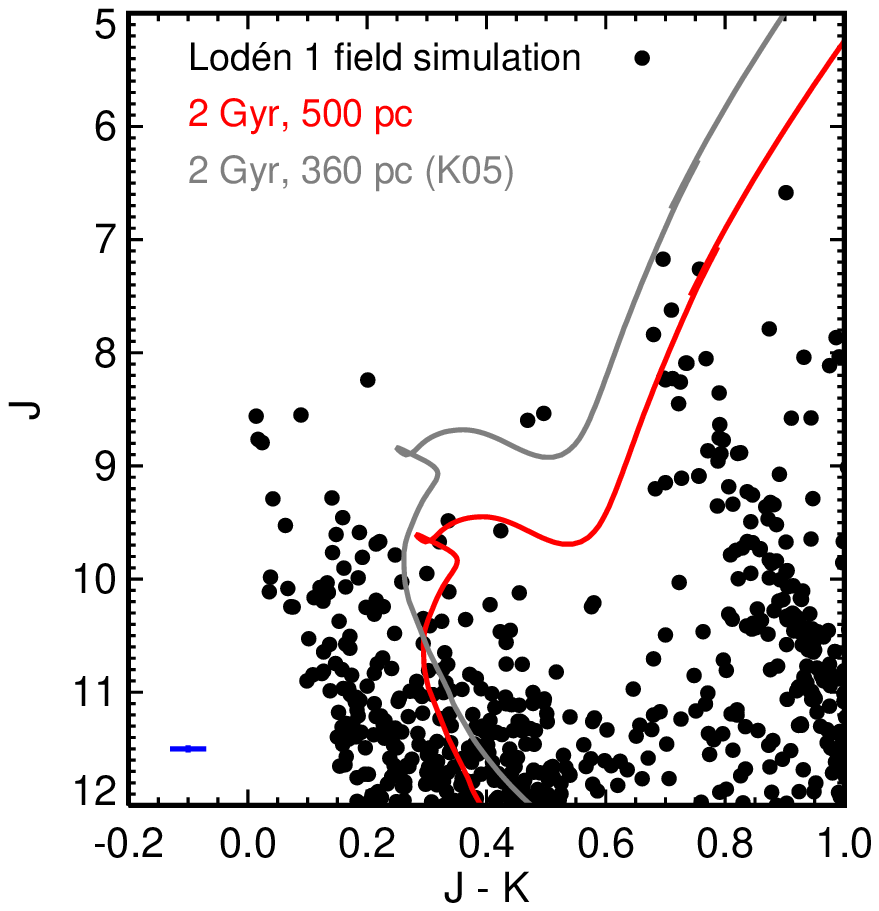}}\\
    \caption{Proper motion and color-magnitude diagrams for \loden\ 1 and representative control fields, 
    similar to Figure \ref{f:r147}.
    Left panels show all stars brighter than $J < 12$ within the innermost 30 arcmin 
    of \loden\ 1. 
    Middle panels show a neighboring field of equivalent size and magnitude cutoff
    at fixed Galactic latitude and shifted by $+1$\fdg5 in longitude. 
    Right panels show a simulation of the primary \loden\ 1 field, 
    generated with the Besan\c{c}on population synthesis model for the Milky Way
    using the same location, area, and magnitude selection criteria that 
    defined the primary sample. 
    These data were queried from PPMXL.
    We see no obvious evidence for a cluster within 500 pc, 
    in sharp contrast to the R147 field. 
    \textit{If} a real cluster exists within 500 pc in this field, it must be extremely sparse.
    }
    \label{f:loden1}
\end{center}
\end{figure*}

Let us begin by examining the proper motion distributions for 
the cluster and neighboring control fields and the Galactic model simulation samples, 
which we plot in the top panels of 
Figures \ref{f:r147} and \ref{f:loden1} for R147 and \loden\ 1. 
The typical astrometric error for the PPMXL proper motions in this magnitude range is 5 \mas ; 
for individual candidates (Table \ref{t:mem}), the errors range from 1 to 6 \mas , 
except for \loden\ 1 20 where $\sigma_\mu$ = 15.7 \mas . 
The bottom panels of these figures show 2MASS $J$ versus $J - K$ color--magnitude 
diagrams (CMD) for these samples. 
The typical photometric errors are under 0.03 mag 
(the mean and standard deviation of $\sigma_J$ for the nearest 936 stars to the \loden\ 1 field with $8 < J < 12$ are 
0.024 $\pm$ 0.003 mag).

The proper motion distribution in the R147 field (Figure.~\ref{f:r147}, left) shows 
kinematic clumping clearly distinct from the field, 
which is emphasized by visual comparison to the 
control sample (center) and simulation sample (right). 
The \loden\ 1 field and control samples visually appear quite similar
and show no obvious clumping beyond the field distribution. 

Identification of the R147 membership in the CMD
is greatly facilitated by selecting stars within this kinematic clump, 
but this is not strictly necessary. 
At only 300 pc the R147 upper main sequence and RGB stand out from the 
Galactic field in the CMD. 
We see no obvious overabundance along the \loden\ 1 stellar locus 
defined by the \citet{khar2005} model, nor even at 500 pc. 

\subsection{A Closer Look with Nearest-neighbor Filtering}
If we assume that the control sample accurately represents 
the field star distribution of the target sample 
in terms of kinematics, stellar composition, and richness, 
then a statistical subtraction of the control sample from 
the target field might reveal hidden structure in 
the proper motion and CMDs. 
We perform a simple subtraction by identifying the 
``nearest neighbor'' of a particular star in the control field 
relative to the target sample, 
and removing it from the target list if this neighbor is within some threshold distance
(0.05 mag for CMD subtraction, 5 \mas\ for proper motion subtraction); 
this test and subtraction is carried out on every object in the control sample. 
If an overdensity was masked by the field distribution, 
this subtraction might reveal it 
if the underlying cluster is rich enough 
to survive the subtraction. 

\begin{figure*}\begin{center}
\subfloat[]{\includegraphics[width = 3.0in]{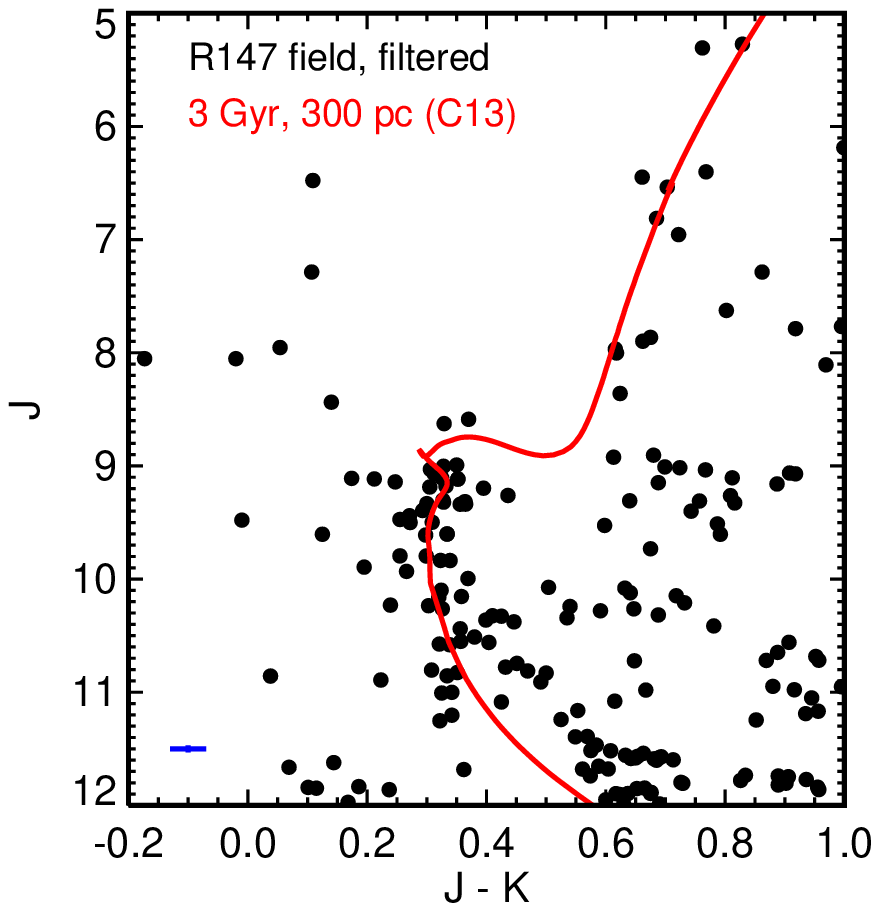}} 
\subfloat[]{\includegraphics[width = 3.0in]{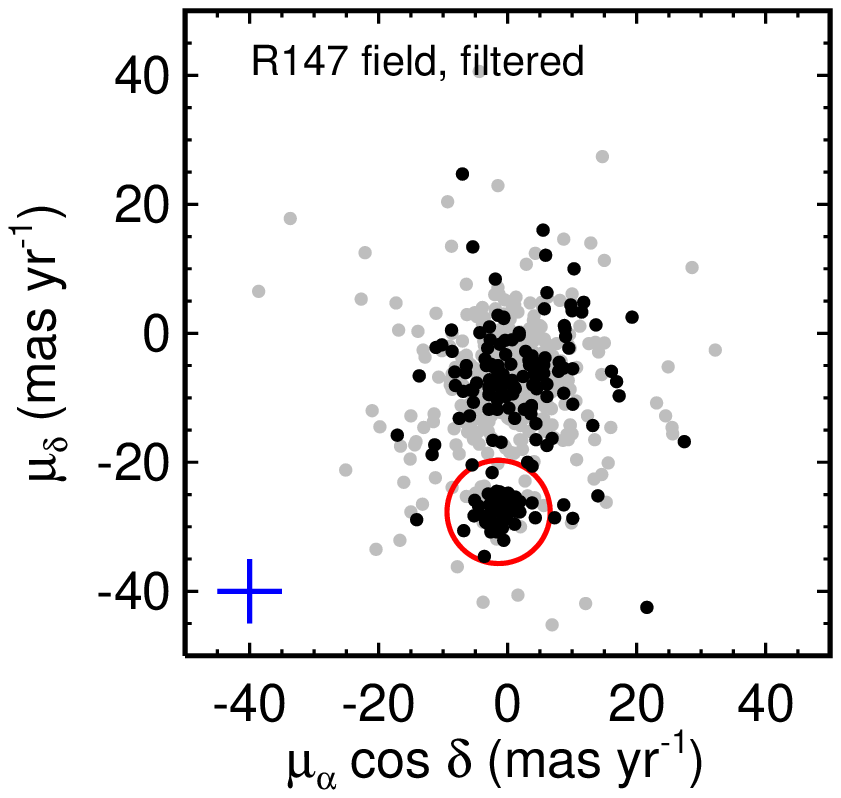}}\\
\subfloat[]{\includegraphics[width = 3.0in]{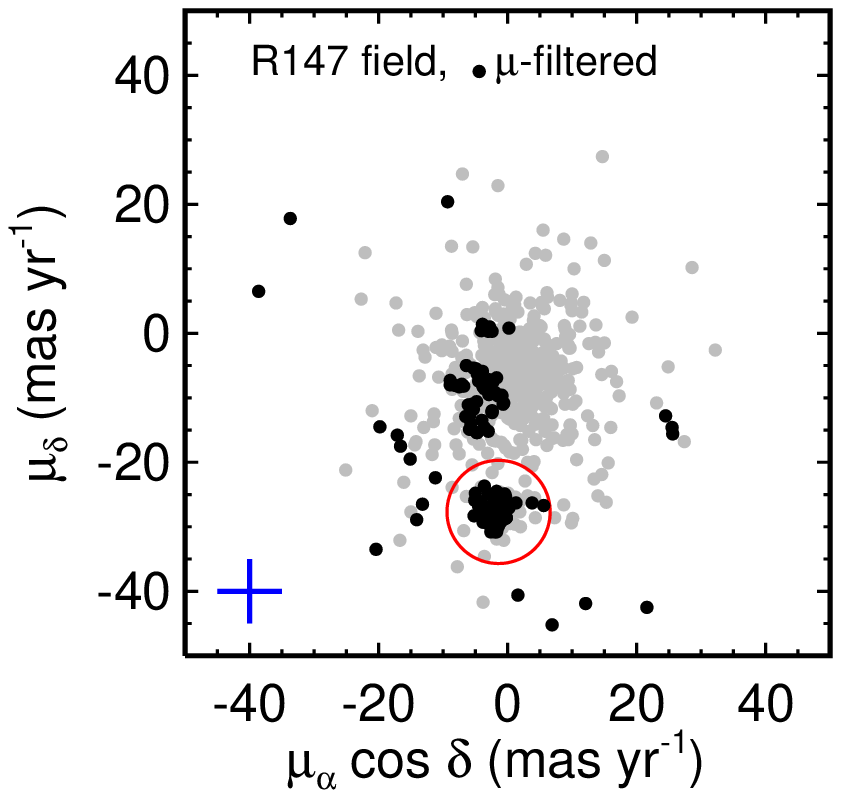}} 
\subfloat[]{\includegraphics[width = 3.0in]{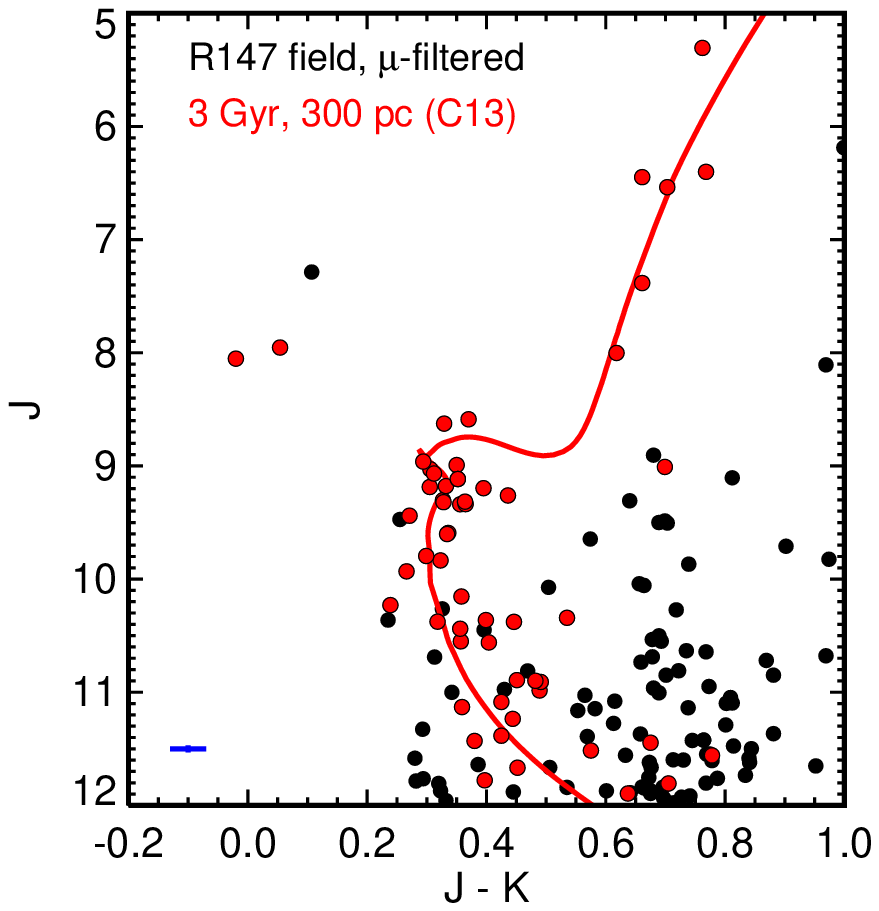}}\\
    \caption{Filtered proper motion and color-magnitude diagrams for R147 using ``nearest neighbors'' subtraction.
    (a) Hundreds of field stars have been removed by filtering the R147 2MASS CMD from Figure \ref{f:r147}-d 
       by removing stars with similar photometry found in the neighboring control field in Figure \ref{f:r147}-e. 
    (b) The proper motion distribution for those stars that survived this particular subtraction 
       highlights how the Galactic field (Fig.~\ref{f:r147}-a, shown in gray) has been substantially rejected by this procedure.
    (c) The proper motion distribution for the R147 field from Figure \ref{f:r147}-a has been filtered by 
       removing stars with similar motion found in the neighboring control field in Figure \ref{f:r147}-b. 
       Since R147's motion is so large compared to the bulk of the field, 
       this technique performs even better than CMD-filtering and leaves few residual stars outside
       of the cluster's proper motion value. 
    (d) The CMD for stars that survived proper motion filtering, referred to as $\mu$-filtered in the figure annotation, 
       with those stars within 7 \kms\ of R147's motion highlighted in red, along with the isochrone model preferred 
       by \citet{Curtis2013}. 
       Most stars in this panel follow the R147 isochrone model, and the $\mu$-filtering revealed a 
       few blue stragglers as well.
    }   
    \label{f:rfilter}
\end{center}
\end{figure*}

We found this simple technique encouraging when applied to the R147 field. 
Filtering the CMD this way successfully 
removed much of the Galactic background, 
and returned a much cleaner and better-defined main sequence 
(Figure.~\ref{f:rfilter}, top left). 
The remaining stars more prominently show the cluster clump 
in the proper motion diagram (PMD, top right): 
originally 72 out of 600 stars fell within 5 \mas\
of the cluster's motion; after CMD-filtering, 
the ratio is up to 55 out of 146. 
Filtering the proper motion diagram (referred to as $\mu$-filtering in the figures) 
left a few residual groupings 
where the control sample did not perfectly subtract out the Galactic field (bottom left), 
but R147 remains the dominant structure in the PMD. 
The CMD for this $\mu$-filtered sample shows a cleaner R147 sequence (bottom right), 
and those within the R147 kinematic clump (circled red)
are highlighted red in the CMD. 

\begin{figure*}\begin{center}
\subfloat[]{\includegraphics[width = 3.0in]{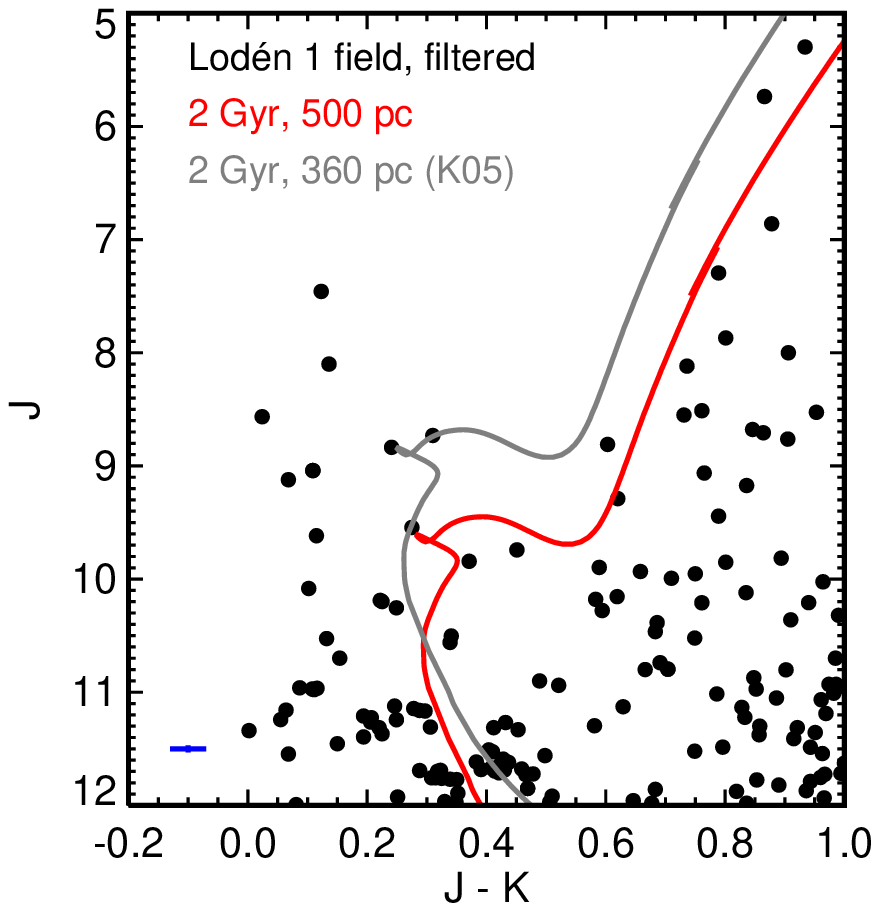}} 
\subfloat[]{\includegraphics[width = 3.0in]{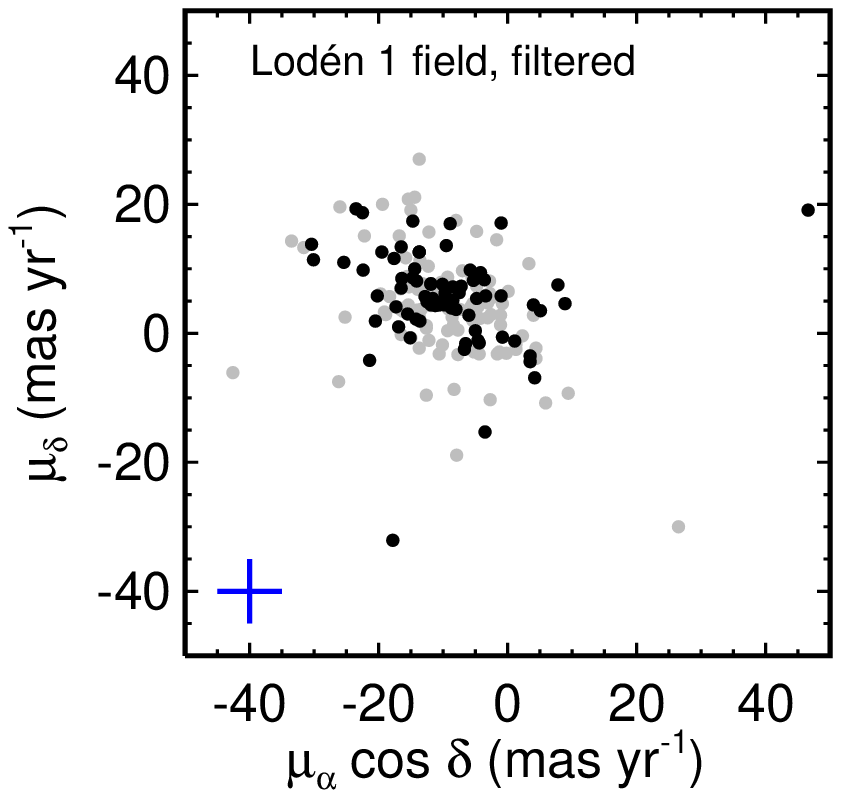}}\\
\subfloat[]{\includegraphics[width = 3.0in]{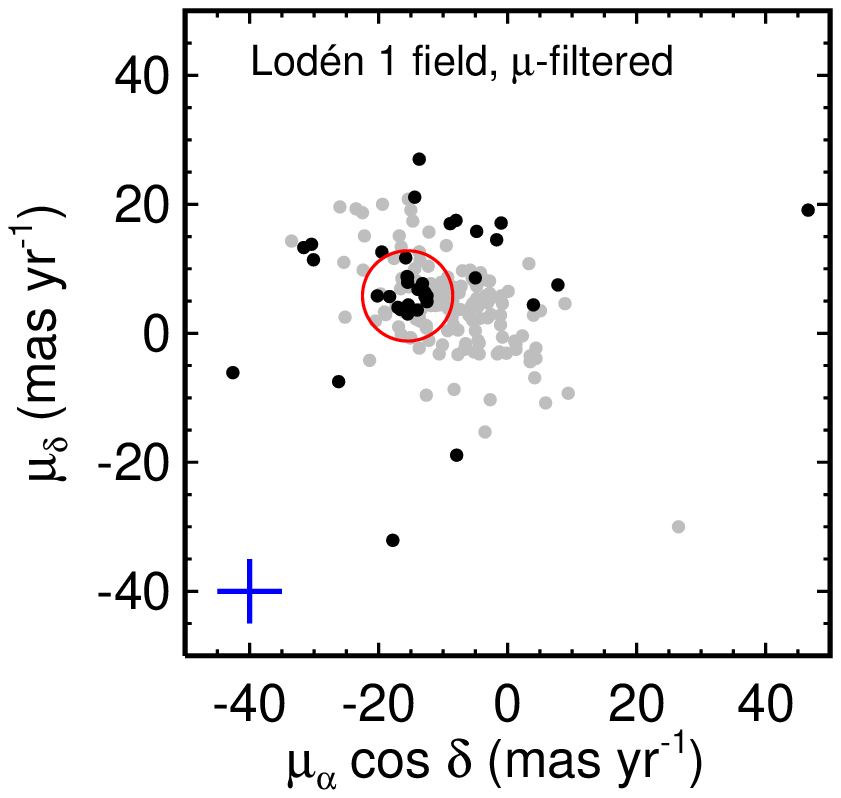}} 
\subfloat[]{\includegraphics[width = 3.0in]{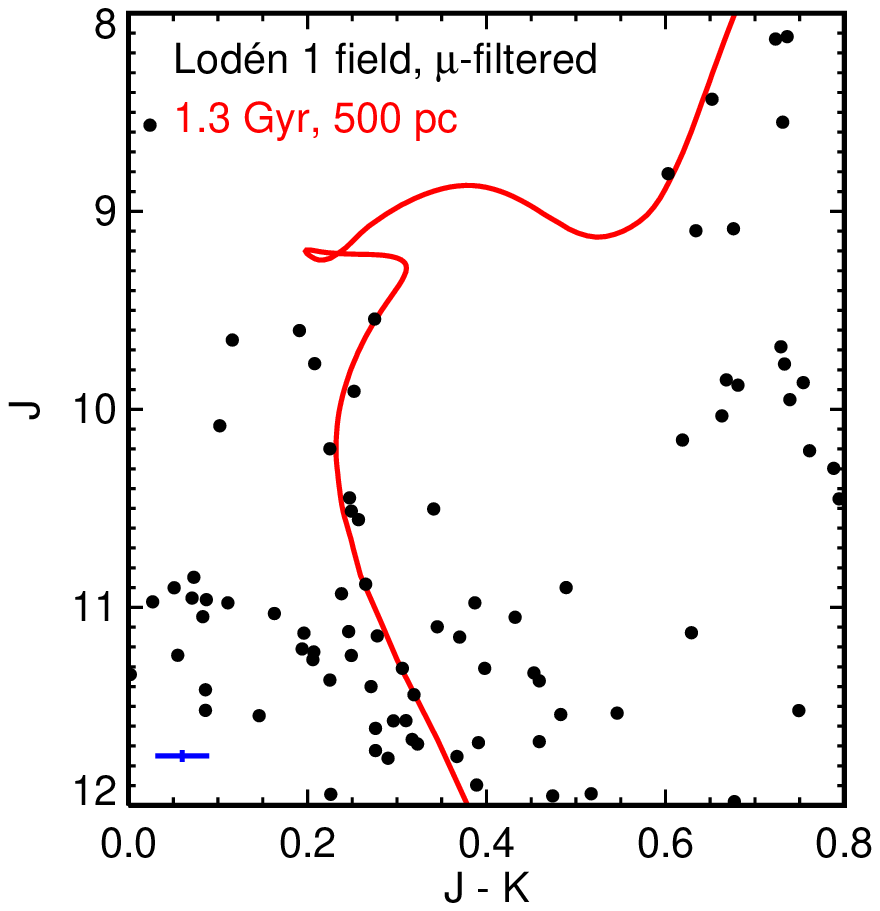}}\\
    \caption{Filtered proper motion and color-magnitude diagrams for \loden\ 1 
    using ``nearest neighbors'' subtraction, similar to Figure \ref{f:rfilter}.
    While filtering the R147 proper motion and color-magnitude diagrams using 
    helped reveal the cluster against the Galactic field, 
    we find no such enhancement for \loden\ 1. 
    (a) The \loden\ 1 CMD from Figure \ref{f:loden1}-d has been filtered by 
       by removing stars with similar photometry to that found in the neighboring control field in Figure \ref{f:loden1}-e. 
       A 2 Gyr Dartmouth isochrone model \citep{dartmouth}, with R147's metallicity and visual extinction assumed
       ([Fe/H] = +0.1, $A_V = 0.25$) and placed at 360 pc is overlaid in gray \citep{khar2005} and at 500 pc in red.
       Only a handful of stars near the 2 Gyr model CMDs survived, and no coeval sequence was revealed. 
    (b) Black points represent those stars that survived this particular subtraction, 
        which trace the overall proper motion distribution for the field shown in gray. 
    (c) The proper motion distribution for the \loden\ 1 field from Figure \ref{f:loden1}-a, shown here in gray, 
       has been filtered by removing stars with similar motion found in the neighboring control field 
       in Figure \ref{f:loden1}-b. 
       The black points reveal a residual clump at $\mu_{\alpha} \cos \delta = -16, \mu_\delta = 6$ \mas 
       and circled red.
    (d) The CMD for all stars in the \loden\ 1 field within 6 \mas\ of this proper motion value, 
       along with a 1.3 Gyr Dartmouth isochrone ([Fe/H] = +0.1) with $A_V = 0.5$ at 510 pc. 
        Nine stars brighter than $J < 11$ appear to follow this coeval sequence, 
        including two on the red giant branch.
        We are curious to see how these stars are actually distributed in space and motion following the 
        first Gaia data release, scheduled for release at the end of northern summer in 2016 (see \S \ref{s:end}), 
        but we are skeptical that these stars are associated. 
        Table \ref{t:mem-jc} provides the 2MASS identities, PPMXL proper motions, and 2MASS photometry for these stars, 
        along with two blue straggler candidates near the putative turnoff and seven additional stars along the isochrone 
        at $J > 11$.}
    \label{f:lfilter}
\end{center}
\end{figure*}

Let us now apply the same filtering technique to the \loden\ 1 field sample, 
using the annulus region as the control sample.
Figure \ref{f:lfilter} shows that the photometric filtering subtracted out most of the stars in the field, 
leaving only a handful of stars brighter than $J < 11$. 
No cluster within at least 500 pc has been revealed by this technique, 
and the motions of these stars trace the proper motion distribution of the field overall.
 
Filtering the \loden\ 1 field based on the proper motion distributions leaves
very few remaining stars, 
likely meaning the control sample accurately represents the target sample. 
We can examine these residuals more closely by selecting all stars in 
the target sample with proper motions within 5 \mas\ of the largest 
residual clump at $\mu_\alpha \cos \delta = -16, \mu_\delta = 6$ \mas .
This is similar to, but not quite the cluster motion detected by \citet{khar2005} 
at $\mu_\alpha \cos \delta = -10.4, \mu_\delta = 4.2$ \mas .
Interestingly, nine stars brighter than $J < 11$ do congregate along an apparently coeval sequence 
at 510 pc. Assuming [Fe/H] = +0.1 (the metallicity of R147), and assuming $A_V = 0.5$ 
from a basic $A_V = 1$ mag kpc$^{-1}$ extinction law \citep{AllenAstroQuant}, 
these nine stars can be fit with a 1.3 Gyr Dartmouth isochrone.
Their 2MASS identities, $(J, K)$ photometry, and PPMXL proper motions are provided in Table \ref{t:mem-jc}, 
along with an additional seven fainter stars that blend into the Galactic background at $J > 11$, 
and two stars that are 0.04--0.08 mag blueward in $J - K$ color 
of the putative main-sequence turnoff, which could be blue straggler candidates.
While it might be tempting to interpret their kinematic and photometric 
association as evidence for an association or cluster remnant, 
we remind the reader that nine stars (or even twenty) 
spread over a 4 pc radius (assuming 500 pc and a 30 arcmin radius) 
cannot remain gravitationally bound for long.\footnote{Assuming 
a generous cluster mass of 20 \msun, and a half-mass radius of 2 pc, 
the virial theorem predicts a velocity dispersion of 65 \mps .}
Even so, at 510 pc and 1--2 Gyr, 
the association would only barely meet our conditions for \loden\ 1 being old, nearby, 
\textit{and useful} as a benchmark system.

Thus, a proper motion and photometric check of the brightest stars in 
the field revealed no compelling evidence for a real cluster within 500 pc.  
Next, we attempt to validate the cluster's existence 
as defined by \citet{loden-mem}.

\begin{deluxetable*}{lccccccc}
\tabletypesize{\scriptsize}
\tablecaption{\loden\ 1 Astrometric-photometric Candidates \label{t:mem-jc}}
\tablewidth{0pt}
\tablehead{
\colhead{2MASS ID} & 
\colhead{$\mu_{\alpha} \cos \delta$} & \colhead{$\mu_{\delta}$} & \colhead{$\sigma_{\mu}$} &
\colhead{$J-K$} & \colhead{$J$} & \colhead{$\sigma_{JK}$} & 
 \colhead{Notes}\\
\colhead{} & \colhead{(\mas)} & \colhead{(\mas)} & \colhead{(\mas)} &
\colhead{} & \colhead{} & \colhead{}}
\startdata
10025969-5556088 & -13.3 & 8.6 & 7.8 & 0.39 & 11.68 & 0.03 &  ... \\
10030013-5546358 & -19.1 & 3.3 & 2.2 & 0.19 &  9.60 & 0.03 &  T \\
10042100-5553597 & -11.6 & 5.4 & 2.5 & 0.27 &  9.54 & 0.03 & \loden 1 9, T \\
10042719-5550584 & -13.4 & 3.7 & 2.8 & 0.65 &  8.43 & 0.04 & \loden 1 7, T \\
10051941-5546252 &  -9.2 & 6.8 & 5.8 & 0.37 & 11.75 & 0.03 & ... \\
10052049-5555298 & -14.6 & 4.5 & 7.8 & 0.39 & 11.90 & 0.05 &   ... \\
10052079-5547037 & -16.5 & 7.0 & 3.8 & 0.60 & 8.81 & 0.05 &  \loden 1 10, T \\
10022779-5544332 & -16.6 & 6.9 & 2.9 & 0.26 & 10.56 & 0.03 &  T \\
10030233-5539181 & -15.3 & 7.2 & 2.7 & 0.25 & 10.45 & 0.03 &  T \\
10030948-5537066 &  -8.9 & 3.1 & 7.8 & 0.38 & 11.95 & 0.03 &  ... \\
10032335-5532196 & -10.0 & 5.7 & 2.4 & 0.21 & 9.77 & 0.03 &   T \\
10040537-5531475 & -12.8 & 5.7 & 2.7 & 0.28 & 11.14 & 0.03 &  T \\
10051003-5534550 & -10.2 & 5.3 & 5.3 & 0.32 & 11.44 & 0.04 &   ... \\
10060918-5533403 & -10.3 & 5.7 & 2.8 & 0.25 & 9.91 & 0.03 &  T \\
10063798-5541203 & -19.7 & 6.1 & 3.4 & 0.25 & 10.51 & 0.03 &   ... \\
10065383-5539223 & -14.4 & 10.0 & 3.9 & 0.31 & 11.31 & 0.03 &  ... \\
10035034-5612332 & -17.3 & 4.1 & 2.7 & 0.23 & 10.20 & 0.03 &  T \\
10060575-5604076 & -10.3 & 7.9 & 3.5 & 0.27 & 10.88 & 0.05 &  T \\
\enddata
\tablecomments{Column Notes: 
 (1) 2MASS identification and J2000 coordinates; 
 (2--3) R.A. and decl. proper motion in \mas\ from PPMXL catalog; 
 (4) proper motion measurement uncertainty in \mas ;
 (5--7) 2MASS $J-K_{S}$ color, $J$ magnitude, and photometric error as quadratic sum of $\sigma_J$ and $\sigma_K$;
 (8) Notes on individual stars: those noted with `T' are included in Tycho-2 and presumably will have high precision 
 proper motions and parallaxes provided in the first \textit{Gaia} data release as part of the Tycho-\textit{Gaia} Astrometric Solution.
}
\end{deluxetable*}

\section{Our targeted survey of Lod\'{e}n 1}
\label{s:survey}
Our primary candidate list is drawn from the 
original discovery paper, \citet{loden-mem}.
His paper listed 29 stars (resolvable with SIMBAD as, e.g., ``\loden 1 1'' through ``\loden 1 29'') 
with $UBV$ photometry and spectral types, 
and coordinates for the brightest 17 stars,\footnote{When we assembled 
our target list, SIMBAD provided coordinates for only these first 17 stars. 
We eventually manually identified the remaining stars with 
an optical image (DSS2R POSS2/UKSTU Red) from the 
Montage web service (http://montage.ipac.caltech.edu),
and confirmed our matching with APASS $BV$ photometry queried from UCAC4 \citep{ucac4}. 
SIMBAD recently incorporated coordinates for these objects from another 
source unknown to us.} along with a finder chart.

\begin{deluxetable*}{llcccccccccccc}
\tabletypesize{\scriptsize}
\tablecaption{\loden\ 1 candidates \label{t:mem}}
\tablewidth{0pt}
\tablehead{
\colhead{ID} & \colhead{2MASS ID} & 
\colhead{$\mu_{\alpha} \cos \delta$} & \colhead{$\mu_{\delta}$} & \colhead{Pr$_{\rm K05}$} &
\colhead{$\mu_{\alpha} \cos \delta$} & \colhead{$\mu_{\delta}$} & 
\colhead{$J-K$} & \colhead{$J$} & 
\colhead{$B-V$} & \colhead{$V$} & 
\colhead{Sflg} & \colhead{RV$_{\rm CCF}$} & \colhead{RV$_{H\alpha}$} \\
\colhead{} & \colhead{} & 
\colhead{(\mas)} & \colhead{(\mas)} & \colhead{} &
\colhead{(\mas)} & \colhead{(\mas)} & 
\colhead{} & \colhead{} & 
\colhead{} & \colhead{} & 
\colhead{} & \colhead{(\kms)} & \colhead{(\kms)}}
\startdata
 1 & 10050462-5546030 &  -9.2 &   3.9 & 98 & -9.8  &   4.4 &  0.93  &  5.30 &  1.60 & 7.98  &  S & 27.7 & 28.5 \\ 
 2 & 10040117-5554295 & -15.7 &   8.8 &  0 & -13.9 &   6.8 & -0.03  &  8.00 & -0.02 & 8.00  & NS & ... & ... \\
 3 & 10033964-5543102 & -29.2 &  13.6 &  0 & -30.4 &  13.8 &  0.14  &  8.10 &  0.19 & 8.57  &  B & ... & 10.7\\
 4 & 10045780-5548020 &  -8.7 &   5.1 & 90 & -11.2 &   4.3 &  0.02  &  8.56 &  0.07 & 8.80  &  F & ... & 21.4 \\
 5 & 10051021-5541135 &  -0.5 & -17.0 &  0 &  -3.5 & -15.3 &  0.31  &  8.73 &  0.43 & 9.71  &  S & -6.1 & ... \\
 6 & 10045082-5550042 & -22.0 &   7.8 &  5 & -20.2 &   5.8 &  0.74  &  8.12 &  1.12 & 10.09 &  S & 15.4 & 15.2 \\
 7 & 10042719-5550584 & -16.2 &   3.3 & 71 & -13.4 &   3.7 &  0.65  &  8.43 &  0.97 & 10.24 &  S & 20.7 & 26.9 \\
 8 & 10053243-5546392 &   3.0 &   8.2 &  0 &   2.8 &   9.9 &  1.27  &  5.67 &  1.42 & 10.10 &  C & ... & ... \\
 9 & 10042100-5553597 & -11.8 &   9.2 & 94 & -11.6 &   5.4 &  0.27  &  9.54 &  0.47 & 10.39 & NS & ... & ... \\
10 & 10052079-5547037 &  -9.0 &  12.1 & 64 & -16.5 &   7.0 &  0.60  &  8.81 &  1.07 & 10.48 &  S & 9.5 & ... \\
11 & 10044032-5554367 &  -4.1 &   6.3 & 38 &  -5.0 &   0.4 &  0.76  &  8.51 &  1.27 & 10.45 &  S & -3.9 & -4.4 \\
12 & 10040233-5550003 &  -5.9 &   9.6 & 42 &  -5.3 &   8.2 &  0.62  &  9.29 &  0.98 & 11.06 &  S & 4.6 & 8.4 \\
13 & 10053487-5540048 & -14.6 &   5.9 & 68 & -10.7 &   7.4 &  0.03  & 10.97 &  0.05 & 11.12 &  F & ... & 16.0 \\
14 & 10034818-5554349 & -14.3 & -36.1 &  0 & -17.8 & -32.1 &  0.45  &  9.74 &  0.78 & 11.22 & NS & ... & ... \\
15 & 10035070-5555134 &  -3.9 &  11.6 &  0 &  -4.2 &   9.4 &  0.15  & 10.70 &  0.30 & 11.35 & S & -51.7 & -50.5 \\
16 & 10034425-5552352 & ...   &  ... & ... & -10.1 &  -1.8 &  0.15  & 11.05 &  0.39 & 11.79 & S & -23.0 & -23.8 \\
17 & 10045726-5555495 &   6.1 &  -6.6 &  1 &   5.9 & -10.8 &  0.23  & 11.22 &  0.24 & 11.87 & S & 65.5 & 60.1 \\
18 & 10050191-5544044 & ...   & ... & ...  & -12.5 &   4.9 &  0.62  & 10.15 &  1.06 & 12.00 & NS & ... & ... \\
19 & 10044239-5547255 &  -7.2 &   8.7 & 83 &  -5.0 &   8.6 &  0.72  &  9.99 &  1.14 & 12.00 &  S & 29.9 & 27.0 \\
20 & 10043846-5554395 & ...   & ...  & ... &  -6.1 &   2.4 &  1.56  &  6.68 &  2.66 & 12.13 & NS & ... & ... \\
21 & 10041815-5546560 & ...   & ...  & ... &  -9.3 &   8.7 &  0.27  & 11.40 &  0.44 & 12.28 & NS & ... & ...  \\
22 & 10050360-5550468 &  -8.0 &   5.8 & 91 &  -6.1 &   2.4 &  0.15  & 11.55 &  0.35 & 12.27 &  S & -6.5 & -6.4 \\
23 & 10041450-5545250 & ...   & ...  & ... & -12.3 &  10.4 &  0.40  & 11.31 &  0.58 & 12.34 & NS & ... & ... \\
24 & 10043620-5542258 & -5.3 &   -1.5 & 70 & -11.6 &  -5.3 &  0.14  & 12.02 &  0.36 & 12.38 & NS & ... & ... \\
25 & 10033695-5552040 & ...   & ...  & ... &  -5.7 &   4.1 &  0.09  & 11.98 &  0.22 & 12.53 & NS & ... & ... \\
26 & 10043076-5544447 & -12.5 &  11.9 & 68 &  -9.1 &  12.8 &  0.15  & 12.07 &  0.32 & 12.69 & NS & ... & ... \\
27 & 10050438-5553591 & ...   & ...  & ... &  -1.1 &   2.8 &  0.38  & 11.49 &  0.58 & 12.70 & NS & ... & ... \\
28 & 10050455-5549545 & ...   & ...  & ... &  -0.8 &  -0.6 &  0.32  & 11.70 &  0.54 & 12.69 & NS & ... & ... \\
29 & 10050445-5554308 & ...   & ...  & ... & -15.0 &  19.1 &  0.33  & 11.81 &  0.57 & 12.97 & NS & ... & ... \\
K09 & 10044926-5537533 &  -9.6 &  12.5 & 43 & -10.1 &   7.6 &  0.85  &  8.68 &  1.39 & 11.07 &  S & 32.5 & 39.4 \\ 
K18 & 10053859-5550335 & -17.0 &  4.8 & 79 & -12.8 &   6.3 &  0.74  &  9.95 &  1.18 & 12.04 &  S & 12.0 & 15.3 \\ 
K19 & 10054220-5553008 & -20.7 & -4.4 & 10 & -21.4 &  -4.2 &  0.22  & 10.19 &  0.38 & 10.98 &  B & ... & 7.9 \\ 
K20 & 10055059-5542335 & -2.6 &   4.3 & 54 & -12.2 &  -1.1 &  0.21  & 11.57 &  0.33 & 12.19 &  F & ... & 1.9 \\ 
K21 & 10055385-5543411 & -18.8 & 12.1 & 47 & -15.8 &  11.7 &  0.68  &  9.88 &  1.03 & 11.81 &  S & 12.3 & 16.4 \\ 
K22 & 10055895-5542015 & -12.2 &  1.4 & 91 & -13.6 &   1.9 &  0.10  & 10.08 &  0.18 & 10.50 &  F & ... & 22.7 \\ 
K23 & 10060137-5547231 & -12.0 & -0.7 & 65 & -11.9 &   4.3 & -0.02  & 10.39 &  0.03 & 10.56 &  F & ... & 11.7 \\ 
K24 & 10060328-5553127 &  -4.4 &  2.5 & 82 &   3.5 &  -4.4 &  0.34  & 11.77 &  0.59 & 12.90 &  S & 8.1  & 15.2 \\ 
K25 & 10061065-5547157 &   1.2 & 11.2 & 37 &   3.3 &  10.8 &  0.76  &  9.61 &  1.15 & 11.71 &  S & -16.1 & -7.9 \\ 
\enddata
\tablecomments{(1) Identification number: 1--29 are the original object numbers from \citet{loden-mem}, stars with ``K'' IDs were identified by \citet{khar2005} as members with kinematic probability $P_{\rm kin} > 0$, the ``K'' notation is our own for easy reference within this paper and we encourage readers to use the 2MASS IDs elsewhere (2) 2MASS ID and coordinates (3, 4) Tycho-2 proper motions (5) \citet{khar2005} kinematic membership probabilities (6, 7) PPMXL proper motion, uncertainties are $\approx$5 \mas , except for \loden\ 1 20 where $\sigma_\mu = 15.7$ \mas (8, 9) 2MASS $J-K_{S}$ color, $J$ magnitude, typical error is $\sigma_{J,K} < 0.04$ (10, 11) $B - V$ color, $V$ magnitude, taken from Tycho-2 \citep{tycho2} for $V < 10.5$ and APASS from UCAC4 \citep{ucac4} otherwise (12) Spectrum description flag: ``S'' = sharp-lined spectrum, ``NS''---no spectrum, ``F''---featureless spectrum with \halpha, ``B''---broad lines, ``C''---Carbon star (13) Radial velocity (\kms) measured in this work via cross-correlation function (CCF), averaged over the 3 spectral segments.  Measurement uncertainties are determined to be $\sigma_{\rm RV} = 3$ \kms\ from analysis of standard stars (14) Radial velocity (\kms) measured in this work, via \halpha\ profile fitting.Measurement uncertainties are determined to be $\sigma_{\rm RV} = 5$ \kms\ from analysis of standard stars. Analyses in this paper use RV from column 14, except for ``F'' spectra where we use RV$_{H\alpha}$\ in column 15, and ``B'' spectra where we use the average from columns 14 and 15. }
\end{deluxetable*}


Table \ref{t:mem} provides 2MASS identifications,
proper motions, optical and NIR photometry, 
and the radial velocities we measure for this targeted sample. 
We supplemented \loden 's list with stars with non-zero 
membership probabilities according to \citet{khar2005}.  
This list can be obtained via VizieR by querying 
the Astrophysical supplements to the ASCC-2.5 catalog \citep{khar2004}\footnote{VizieR table /AN/325/740/csoca} 
for objects with $P_{\rm kin} > 0$ and cluster sequential number, Seq = 236.
These stars are labeled ``K\#'', where the ID number (\#) is taken from the resulting VizieR table entry.

\subsection{Radial velocity survey with SALT}
\loden\ 1 is a Southern object, with declination approximately -55\degree. 
We conducted our radial velocity survey with 
RSS \citep{rss1,rss2,rss3} 
on SALT
\citep{salt1,salt2},
at the Sutherland Observatory in South Africa.
The SALT primary mirror spans 11 m, and is composed of 91 individual 1 m hexagonal mirrors. 
We opted for the pg2300 grating, at angle 46.625 degrees, with a $0.6''$ slit, 
which delivered a spectral resolution of $R \approx 10,000$ 
across 5850--6700 \AA, which ranges between the Na D doublet and \halpha. 
The RSS detector is comprised of 3 CCDs, 
with gaps that span 15 \AA\ in the extracted spectra. 
Figure \ref{f:spec} shows a sample reduced long-slit spectrum, 
which highlights these chip gaps. 
We planned our observations with the RSS Simulator,\footnote{Available at http://astronomers.salt.ac.za/software/\#RSS}
and allocated time to achieve signal-to-noise ratios, $S/N \sim$100. 
For reference, a 120 s exposure of a $V = 10$ star reaches $S/N = 120$. 
Including overhead, the total charged queue time is 851 s.

SALT is a fixed altitude, queue-scheduled telescope, 
similar in design to the Hobby--Eberly Telescope \citep[HET,][]{HET},
and can track objects at most accessible declinations for about one hour at a time. 
Given this pointing constraint, weather, the lunar cycle (this was a ``bright time'' program),
and the 10 minute overhead for target acquisition and instrument configuration, 
our observations were spread out between 2013 May and 2014 January, 
under programs 2013-1-HET-005 
and 2013-2-HET-003.

\subsection{Reducing SALT RSS spectra}
SALT staff astronomers perform the basic CCD processing with the PyRAF\footnote{
    IRAF is distributed by the National Optical Astronomy Observatories,
    which are operated by the Association of Universities for Research
    in Astronomy, Inc., under cooperative agreement with the National
    Science Foundation.
    PyRAF is a product of the Space Telescope Science Institute, 
    which is operated by AURA for NASA.}-based 
data reduction package, PySALT \citep{pysalt},\footnote{http://pysalt.salt.ac.za/}
and the raw and reduced data become available to principal investigators for 
download the following day.\footnote{Nightly observing logs are conveniently 
posted each morning at http://saltastro.blogspot.com}
This pipeline applies gain, overscan and distortion corrections, 
stitches together the images from the 3 CCDs, and converts the data to FITS format. 

We accepted the PySALT pre-processed images as-is 
and completed the spectral extraction and calibration with PyRAF 
using standard commands and procedures 
(e.g., cosmic-ray removal with {\tt xzap}, 
background fitting and one-dimensional spectrum extraction with
{\tt apall} in the {\tt noao.twodspec.apextract} package, 
wavelength calibration with {\tt identify} and {\tt reidentify} tasks 
in the {\tt noao.onedspec} package, 
and dispersion correction with {\tt dispcor} in the same package). 
Every spectrum had a defect at pixel 432, so we set this bad pixel to 
the average value of the neighboring pixels. 

We opted to observe multiple standard stars with known and stable radial velocities 
\citep{carly}, followed by exposures of both an arc lamp 
(typically an argon and/or neon lamp) for 
wavelength calibration, 
so that we could establish the accuracy and precision of the resulting 
wavelength solutions. 
The arc lamp was taken immediately preceding or following the science observation within a minute, 
before the SALT queue observers moved the telescope to their next target. 
Each target was treated as a separate observing block in the queue, 
and was allocated its own calibration lamp exposure. 
The SALT website conveniently provides arc line atlases for each instrument setting.\footnote{http://pysalt.salt.ac.za/lineatlas/lineatlas.html}

\begin{figure}
    \centering
    \includegraphics[scale=0.5]{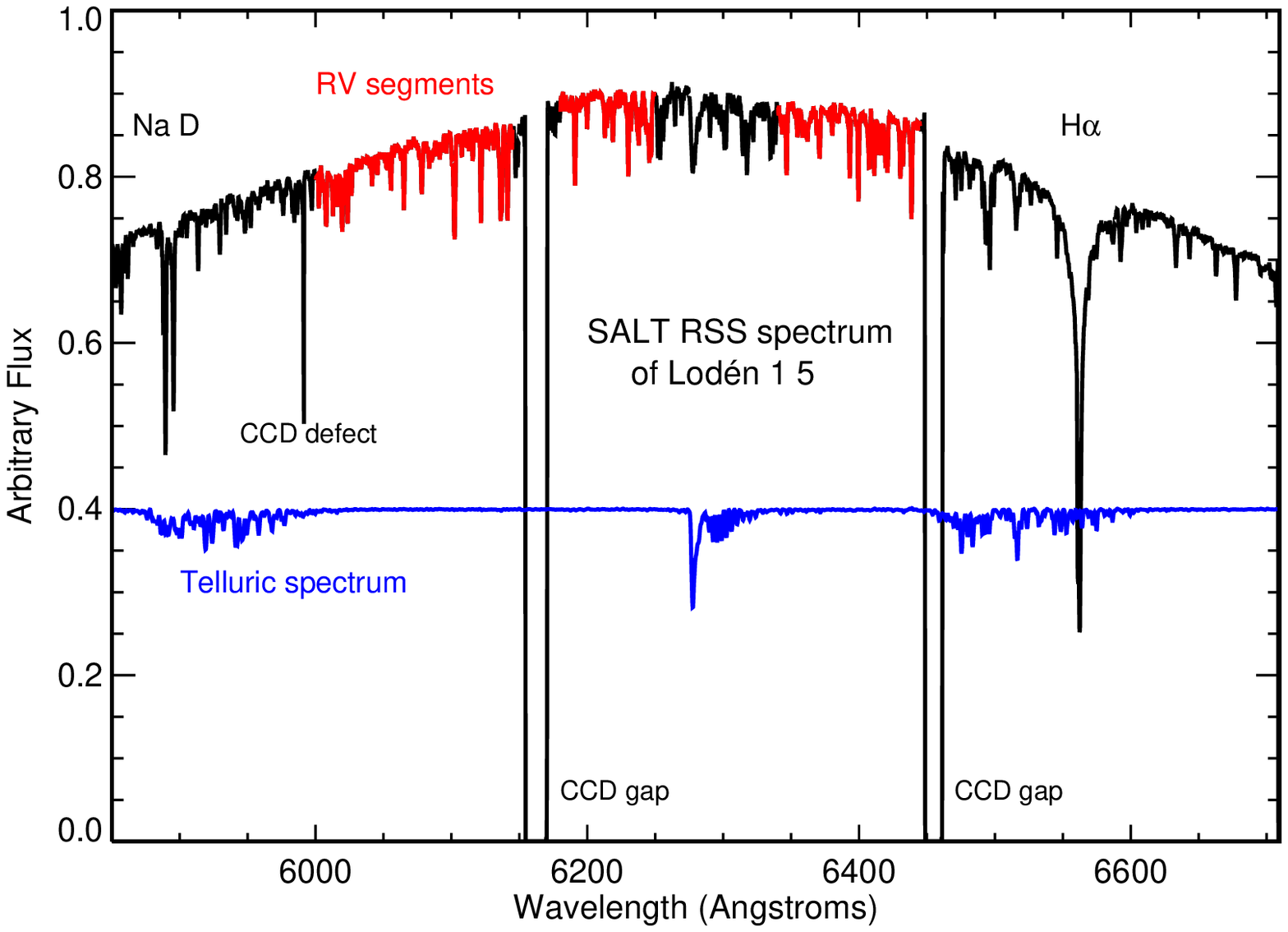}
    \caption{SALT RSS spectrum of \loden\ 1 5. 
        This example spectrum ($R \approx 10,000$) illustrates the RSS CCD gaps and chip defect. 
        We include the \citet{wallace2011} telluric spectrum smoothed to our instrument resolution (blue),
        and highlight three telluric-free segments used to calculate radial velocities (red). 
    }
    \label{f:spec}
\end{figure}

The sample spectrum in Figure \ref{f:spec} shows the defect at pixel 432 prior 
to correction. The middle segment appears to suffer from a minor gain 
error, although this has no impact on the radial velocities we
measured from each CCD segment.
The \citet{wallace2011} telluric spectrum is shown in blue.

\subsection{Radial velocity calibration}
We calculated RVs with {\tt xcsao} ({\tt rvsao}), 
and applied barycentric corrections calculated with {\tt bcvcorr}, 
to account for Earth's orbital motion. 
RVs were also measured separately in IDL with our own procedures. 
In IDL, we binned the spectra logarithmically in wavelength, 
cross-correlated each spectrum against a reference star to find relative velocities \citep{rvs}, 
then applied the absolute RV of that reference, taken from \citet{carly}.   
Barycentric corrections were calculated in IDL with {\tt baryvel}.\footnote{Available 
from NASA Goddard \citep{idlastro}, http://idlastro.gsfc.nasa.gov/}

Figure \ref{f:spec} shows considerable atmospheric absorption across our spectral range, 
with three relatively telluric-free regions highlighted in red. 
We measured radial velocities in each segment separately and averaged the result. 
This allowed us to (1) effectively mitigate telluric contamination by avoiding problematic spectral regions, 
and (2) verify negligible systematic errors in the wavelength solution due to the CCD stitching procedure. 

We observed seven standard stars with stable radial velocities according to 
\citet[][rms $<$0.15 \kms\ for at least two observations, 
except for HD 43745, which was only observed once in that work]{carly} 
in order to assess the accuracy and precision of the RVs we derived 
with our instrumental configuration and analysis procedure.
These observations are listed in Table \ref{t:stand}, 
along with the spectral types from SIMBAD, and absolute RVs, 
date of observation, 
and the difference between our measured RV and the absolute RV.

The RVs computed in the three wavelength segments agreed to within $\sim$5 \kms, 
and were averaged together to produce our reported values. 
These RVs are consistent to within 3--4 \kms\ of the 
absolute RV values from \citet{carly}. 
HD 35974 was observed on 2013 December 21 
and one month later on 2014 January 19, 
and the RVs differ by 0.4 \kms.
Considering the distribution of $\Delta$ RV, 
we assume a typical measurement error of 3 \kms, 
and will search for \loden\ 1 candidates 
with velocities clustering within 6 \kms\ of each other.
This measurement error is on par with the accuracy of the wavelength solution achievable with this 
instrument configuration, which we found to be $\approx$1.3 \kms, 
and is consistent with the accuracy measured by SALT astronomers 
with the lower resolution gratings (10 \kms\ for PG900, 5 \kms\ for PG1800).\footnote{Information regarding 
performance of other gratings conveyed over e-mail from SALT staff.}

\subsection{Rapidly rotating stars}
The spectra of seven candidate stars appear smooth and nearly featureless, 
with the exception of Na D (possibly interstellar), 
and \halpha. 
The $B - V$ colors of these stars are consistent with 
$\teff\ > 8000$ K, 
and their \halpha\ line profiles appear rotationally broadened by 
$50-200$ \kms, which is common for A stars.
As \citet{Becker2015} recently demonstrated, 
even a few strong features in rapidly rotating B and A stars 
can yield absolute RVs with precisions ranging from 
0.5 to 2 \kms. 

We measured \halpha\ centroids by fitting Gaussian and Voigt functions 
to the line profiles, 
derived RV$_{\halpha}$ for six of our standard stars,
and found values compared to the literature values and 
the accuracy and precision from the cross-correlation method using 
different spectral segments. 
The line core for the seventh standard, HD 71334, shows \halpha\ in emission and 
off-center relative to the wings, and we ignored it for this purpose. 
The \halpha\ RV for HD 68978 is also off by 10 \kms  
due to a poor wavelength calibration. 
Still, the performance of the remaining five standards supports our 
use of \halpha\ for radial velocity measurements at a precision and accuracy of 
$\sim$5 \kms .

\begin{deluxetable}{llcccc}
\tabletypesize{\scriptsize}
\tablecaption{Radial velocity standard stars \label{t:stand}}
\tablewidth{0pt}
\tablehead{
\colhead{HD name} & \colhead{SpT} & \colhead{RV} & 
\colhead{Date} & 
\colhead{$\Delta$ RV$_{\rm CCF}$} & \colhead{$\Delta$ RV$_{\halpha}$} \\
\colhead{} & \colhead{} & \colhead{(\kms)} & \colhead{} & 
\colhead{(\kms)}  & \colhead{(\kms)}}

\startdata
19467                  & G3   & 7.00 & 2014 Dec 01 & 0.9  & 0.6 \\
43745\tablenotemark{a} & F8.5 & -2.4 & 2014 Jan 18 &  ...   &  2.3 \\ 
35974                  & G1   & 76.5 & 2013 Dec 21 & 2.8 &  2.7 \\
                       &      &      & 2014 Jan 19 &  3.2 &  5.6 \\
45184                  & G1.5 & -3.9 & 2014 Jan 19 & -2.1 & -2.7 \\
63754                  & G0   & 45.0 & 2014 Jan 18 & 4.4 & -0.2 \\
68978                  & G0.5 & 51.7 & 2014 Jan 18 &  3.0 & 9.9 \\
71334                  & G2.5 & 17.4 & 2014 Jan 18 & -0.4 & 

...\tablenotemark{b} \\
\enddata
\tablecomments{(1) HD Name,  (2) spectral type from SIMBAD,
    (3) radial velocity from \citet{carly},
    (4) date of observation, 
    (5) difference between our measured radial velocity via 
    cross-correlation and value cited in Column 3,
    (6) difference between our measured radial velocity via 
    \halpha\ and value cited in Column 3}

\tablenotetext{a}{Our reference---all RVs computed with respect to this spectrum.}
\tablenotetext{b}{\halpha\ line core emission reversal, 
unable to accurately model with a simple Voigt profile.}
\end{deluxetable}
\subsection{Radial velocity of candidate members}
\begin{figure}
    \centering
    \includegraphics[width=0.5\textwidth]{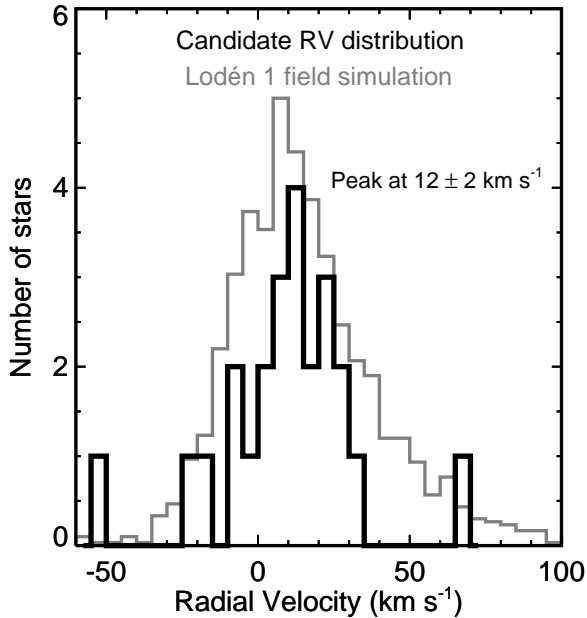}
    \caption{Radial velocities for 24 candidates measured from SALT RSS spectra. 
        Table \ref{t:mem} provides measurement values, quality details, and uncertainty estimates.
        The bin width is 5 \kms\ to account for the typical measurement uncertainty.
        The gray histogram shows the RV distribution for the Galactic field that is expected from the 
        simulated control field, normalized to a value of five.
        Seven stars have RVs within 6 \kms\ of the distribution peak at $12 \pm 2$ \kms: 
        \loden\ 1 3, 6, 10, 13; K18, K21, and K24. 
        Five of these stars show similar proper motion; however,  
        the proper motions for \loden\ 1 3 and K24 are too discrepant from the other five ($\sim 16, 22$ \mas ) 
        to warrant further attention.  
        The remaining five stars are highlighted as orange squares in the CMDs in Figure \ref{f:cmd}---they do not follow a coeval sequence, 
        we therefore do not suspect physical association.
        \label{f:rv}}
\end{figure}

We observed 25 unique candidates in total, and 14 candidates were not observed. 
\loden\ 1 22 was observed twice, on 2013 June 18 and 2014 January 16, 
and the cross-correlation function (CCF) velocities differ negligibly, 
and the \halpha\ velocities differ by $\approx$2 \kms,
validating our precision estimate.
The RVs for those 16 stars from both CCF and \halpha\ methods 
are typically consistent to within $\sim$4 \kms\ given by the standard deviation; 
however, there are a few outliers at $\Delta {\rm RV} = 7-8$ \kms . 
Table \ref{t:mem} presents these radial velocities, 
which are plotted as a histogram in Figure \ref{f:rv}. 
The analysis of these measurements will continue in Section \ref{s:nor}.

\section{Is Lod\'{e}n 1 an old nearby star cluster?}
\label{s:nope}
We can photometrically test the candidate list for age and proximity.
Assuming a substantial fraction of the proposed \loden\ 1 cluster, 
as originally defined by \citet{loden-mem}
and expanded on by \citet{khar2005}, 
constitutes a real open cluster, 
the stars should congregate on a CMD 
within a stellar locus described by a single isochrone model. 

\subsection{Lod\'{e}n 1 Is Neither Old, Nor Nearby...}
\begin{figure*}\begin{center}
    \plottwo{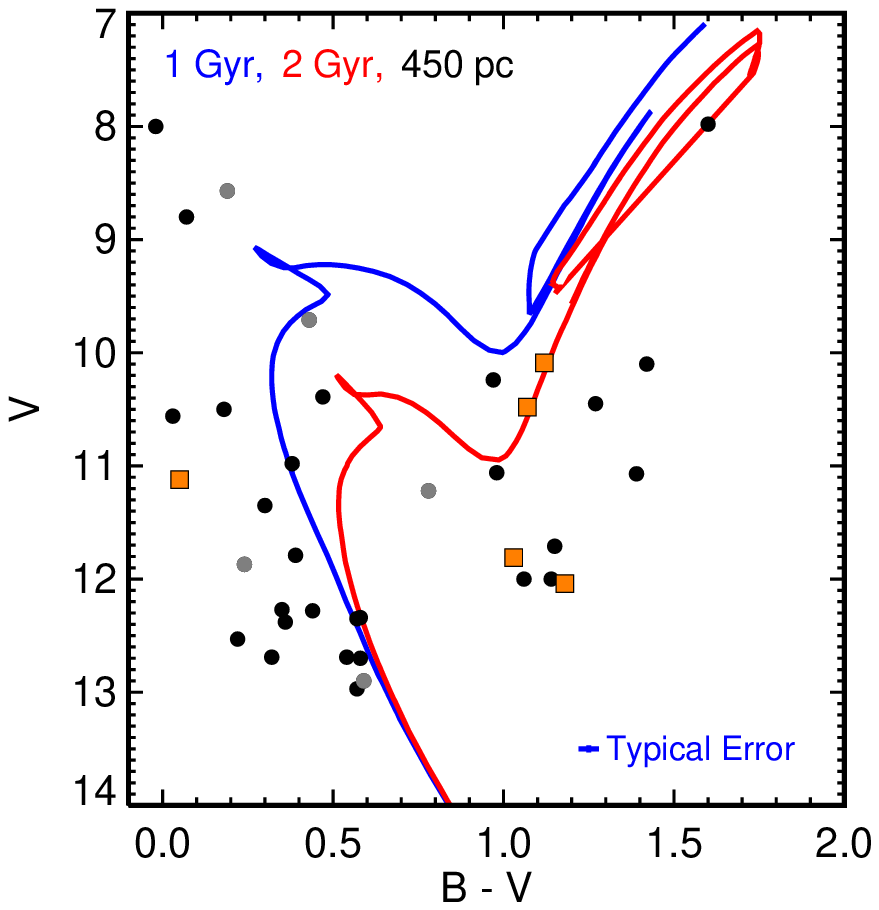}{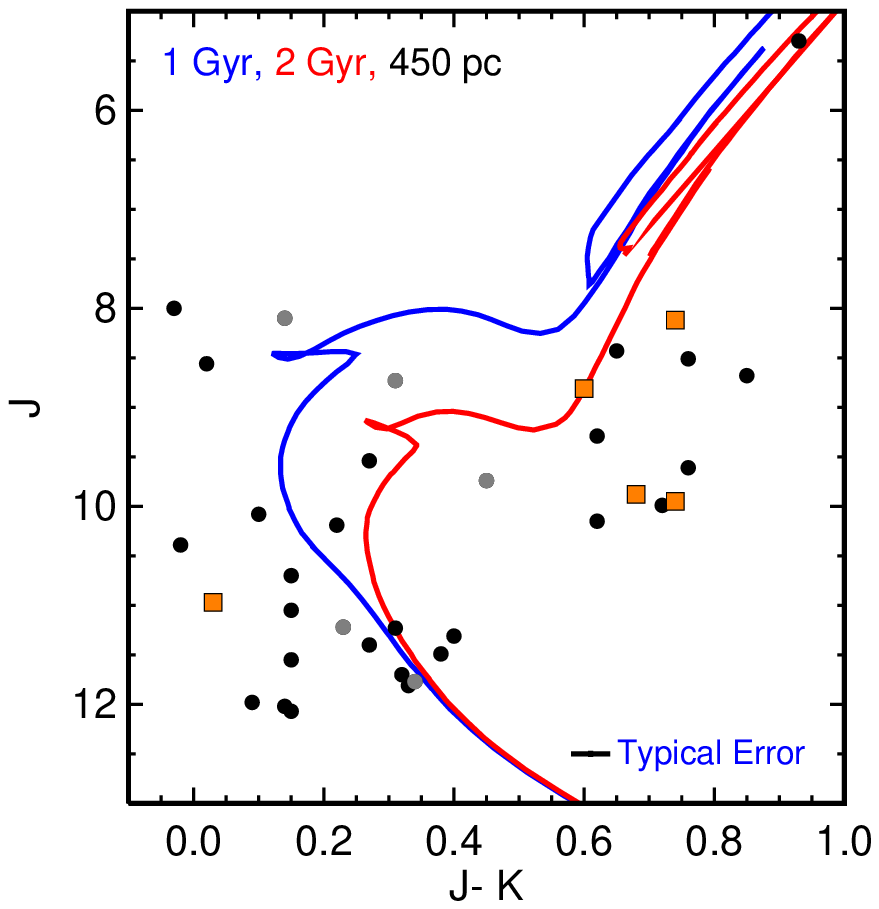}
    \plottwo{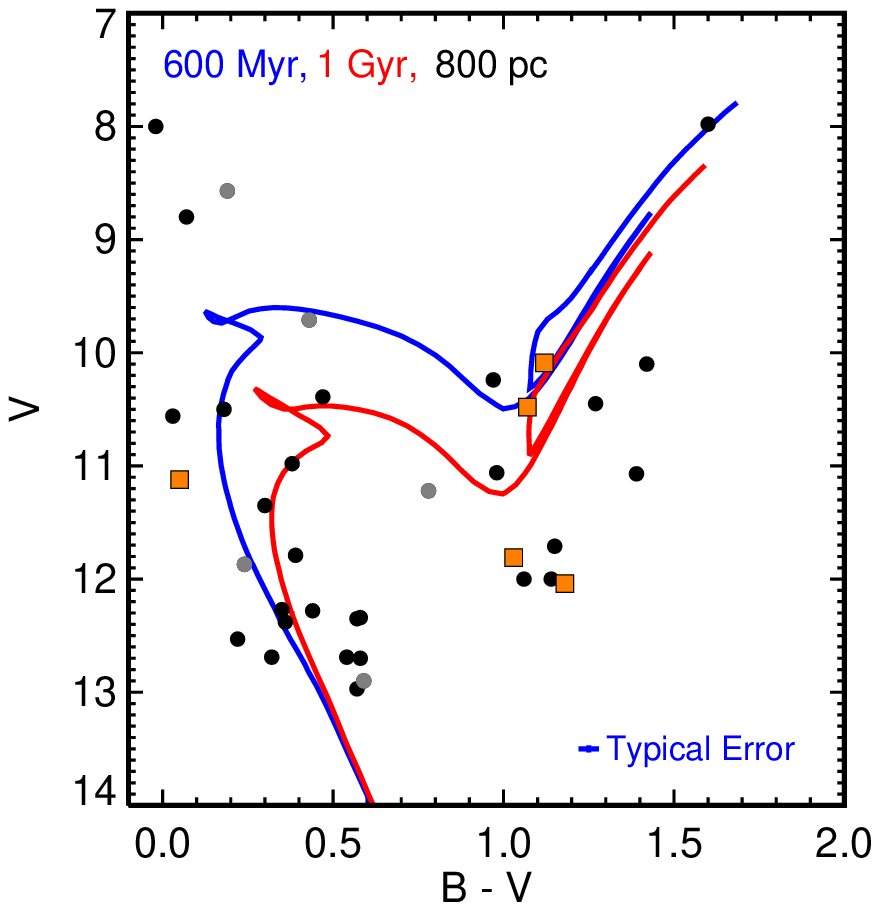}{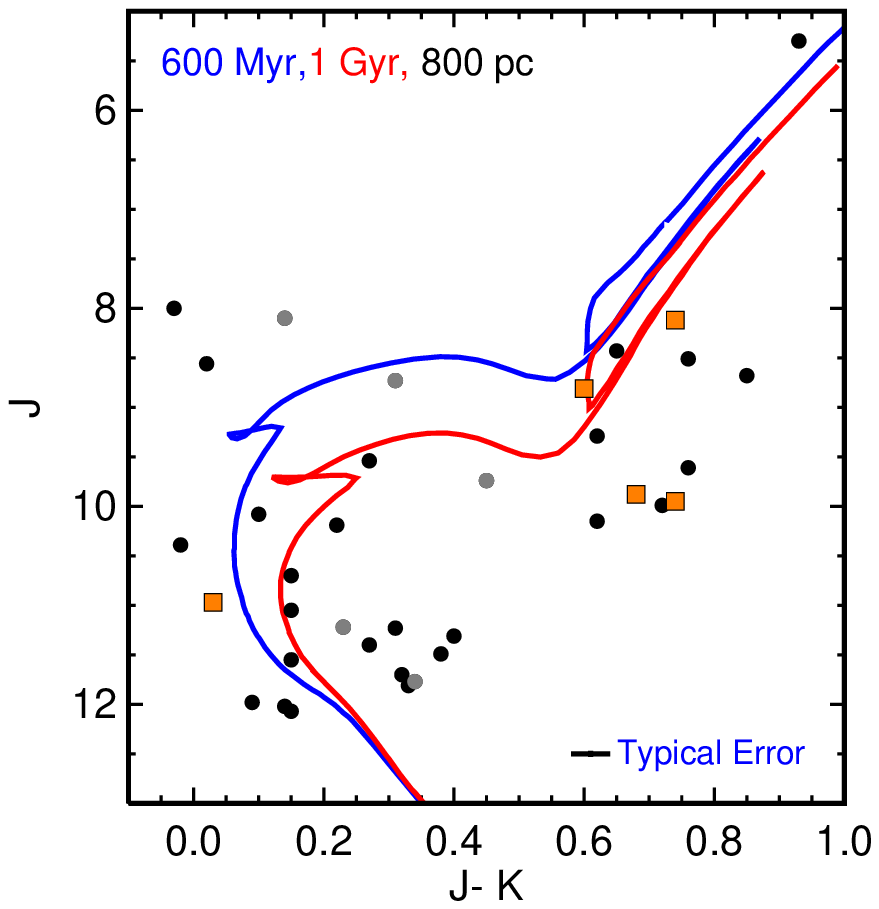}
    \caption{Optical and NIR CMDs for the putative \loden\ 1 cluster. 
        All proposed members listed in Table \ref{t:mem} are plotted.
        Five stars that are clear proper motion outliers have been shaded gray
        and should be disregarded.
        Five other stars with RVs within 6 \kms\ of the 
        distribution peak are marked by orange squares (Figure.~\ref{f:rv}).
        Top panels show 1 and 2 Gyr Solar metallicity PARSEC isochrones \citep{parsec} placed at 450 pc. 
        Bottom panels show 600 Myr and 1 Gyr isochrones placed at 800 pc. 
        All isochrone models include a 1 mag kpc$^{-1}$ visual extinction and $R_V = 3.1$ reddening laws. 
        The top panels demonstrate that the majority of these stars cannot be 
        bound in an intermediate-aged (1--2 Gyr) cluster within 500 pc, 
        because most candidates are located at fainter magnitudes and are
        therefore more distant.
        Given the prevalence of hot A stars, 
        the bottom panels illustrate a representative age of less than 1 Gyr, 
        existing at nearly 1 kpc.
        The combination of distant A stars and red giants is 
        entirely consistent with a Galactic plane field star population. 
        As for the five stars with RVs within 6 \kms\ of the RV distribution peak, 
        they clearly do no follow a standard coeval sequence and we conclude that they are unrelated.
        \label{f:cmd}}
\end{center}\end{figure*}

Figure \ref{f:cmd} plots optical and NIR CMDs for the candidates 
(list and photometry in Table \ref{t:mem}) and we 
test for proximity and age with solar metallicity 
PARSEC isochrone models \citep{parsec}.\footnote{Queried from web service: http://stev.oapd.inaf.it/cgi-bin/cmd}
We assume a 1 mag kpc$^{-1}$ foreground extinction law (e.g., $A_V = 0.5$ at 500 pc) 
due to \loden\ 1's location in the Galactic plane. 
The top panel demonstrates that the majority of these stars cannot be 
bound within an intermediate-aged (1--2 Gyr) cluster within 500 pc. 
The CMDs do not show candidates congregating near the main-sequence turnoff or RGB, 
and the majority of candidates are fainter than these models.
\loden\ 1 is therefore not nearby ($d > 500$ pc).

The swath of hot A stars in the lower left quadrants 
indicate an age of less than 1 Gyr, and this subpopulation 
appears to be constrained within 500 Myr to 1 Gyr, 
as shown by the models plotted in the lower panels of Figure \ref{f:cmd}. 
Fitting these A stars requires a distance more akin to M67's at 800--900 pc.
If these stars do form a cluster with age between 500 Myr and 1 Gyr, 
and a distance of $\sim$800 pc, 
this extremely sparse object would be rather unremarkable
compared to the rich, nearby Hyades and Praesepe and distant M37 clusters. 
The \textit{Kepler} cluster NGC 6811 is also available at 1 Gyr and just beyond 1 kpc.
However, the presence of A and RGB stars at $\sim$1 kpc in the Galactic plane 
is also entirely consistent with the field population.

\subsection{...nor a cluster}\label{s:nor}

Figure \ref{f:rv} shows the RV distribution using RV$_{\rm CCF}$, or RV$_{H\alpha}$ when the former is unavailable.  
The RVs of seven stars are within 6 \kms\ of the distribution peak at 12 $\pm$ 2 \kms :  
\loden\ 1 3, 6, 10, 13; K18, K21, and K24.
The proper motions for \loden\ 1 3 and K24 are too discrepant from this group to warrant further attention.  
The remaining five stars are highlighted as orange squares in the CMDs in Figure \ref{f:cmd}. 
They clearly follow no coeval sequence. 

The data support, at best, five to ten stars that might be common members of an equidistant, common proper motion, coeval sequence. 
If the stars with  $V < 11$ (i.e., subgiants and red giants) were a part of such a sequence, then there would be a larger cluster of stars below the turnoff point between $11 < V < 13$ with slightly lower mass, even for a top-heavy cluster. However, there is a lack of such main-sequence stars, as shown in panel (d) in Figure \ref{f:lfilter}. Therefore, there is no reason to suspect that this handful of stars are, in fact, bound, coeval, or equidistant.

\section{Summary}\label{s:end}
Only two star clusters older than 1 Gyr are known (i.e., proven to exist and well-characterized) 
within 500 pc:  NGC 752 and R147. 
\citet{khar2005} first identified R147 as old and nearby, 
along with NGC 2240 and \loden\ 1. 
R147 turned out to be a real cluster and an important new benchmark, 
and was recently surveyed by NASA's \textit{K2} mission during Campaign 7 
\citep{Howell2014}.\footnote{http://keplerscience.arc.nasa.gov/K2/Fields.shtml\#7}
The potential utility that would come with the discovery of a new 
2 Gyr star cluster at a distance of only 360 pc was sufficiently high that it 
warranted careful inspection and disproof.  

\loden\ 1 was originally classified as a cluster because of a
perceived slight overdensity of bright red and blue stars.
We assembled proper motions and measured radial velocities 
for members proposed by \citet{loden-mem} with SALT RSS (achieved 3 \kms\ RV accuracy with $R = 10,000$ setting). 
We find neither kinematic nor photometric evidence for the existence 
of the cluster known as \loden\ 1,  
and we demonstrate that no $>$1 Gyr cluster exists in this field within 500 pc.
The prevalence of hot A stars implies a young age at a large distance, 
and their existence along with the red giants is consistent with 
a Galactic plane field star population. 

While we cannot definitively rule out the existence of bound clusters of 
\textit{any age and distance} in the direction of \loden\ 1, 
we can rule out an \textit{old, nearby cluster} with more than a few bright members.
Any object older than 1 Gyr and that sparse would have a tough time remaining gravitationally bound in the Galactic plane, 
and such an object would offer little utility to our effort of establishing new benchmark clusters. 

John Herschel identified the cluster we now call Ruprecht 147 based
on a similar slight overdensity of bright stars. 
The majority of these stars actually form an asterism, 
while only one-third of the bright stars are cluster members. 
Without R147's favorable kinematic distinction from the field, 
we would not have found this remarkable cluster sitting on our doorstep. 
Similar objects are likely hidden amongst the Galactic field. 
\textit{Gaia}'s high quality 6D position and kinematic database will 
enable a thorough census within 1 kpc. 

Another cluster's existence was also refuted using an earlier version of the PPMXL astrometric catalog. 
\citet{ngc2451} demonstrated that the cluster known as NGC 2451 did not exist, 
while evidence for two other potential clusters in the same field were offered. 
The \textit{Gaia} team has already demonstrated the potential for cluster 
discovery, confirmation, and characterization with \textit{Gaia}, 
where preliminary data were used to confirm the non-existence of NGC 2451 and definitely prove 
the kinematic and spatial clustering of two other objects, now referred to as NGC 2451A and NGC 2451B.\footnote{http://www.cosmos.esa.int/web/gaia/iow\_20151218 } 
This is a remarkable achievement.
The first \textit{Gaia} data release is scheduled for mid-2016, and ESA plans to 
release full five-parameter astrometric solutions (positions, parallaxes, and proper motions) for all Tycho-2 stars 
exhibiting single-star behavior, 
which includes the majority of the \loden\ 1 candidates.\footnote{The first set of proper motions, columns 3 and 4, 
are drawn from Tycho-2, and designate which stars are likely to be included in the first \textit{Gaia} data release.}
It will be educational to revisit the \loden\ 1 field to see how these stars and others in the field 
are actually distributed in space and motion in that field. 
However, even if \textit{Gaia} finds one or more coeval, 
co-moving stellar populations in this field, they will not be ``\loden\ 1.''

Hopefully, we will find new middle-aged clusters in our neighborhood with \textit{Gaia}.
For now, it appears that R147 and NGC 752 remain the only 
known nearby, middle-aged clusters.

\acknowledgments
Preliminary results were presented by \citet{eunkyuposter}.
This work was supported by the National Science Foundation 
under grant No. AST-1211785,  
and the Center for Exoplanets and Habitable Worlds at Penn State. 
J.C. acknowledges support from the National Science Foundation 
Graduate Research Fellowship Program under NSF grant No. DGE1255832.
Any opinions, findings, and conclusions or recommendations expressed 
in this publication are ours and do not necessarily 
reflect the views of the National Science Foundation.

We appreciate Professor Phil Muirhead and his group members' careful reading of an early draft of this manuscript, 
and we thank the anonymous referee for thorough and thoughtful reports that have significantly improved our paper.

All spectroscopic observations reported in this paper were obtained with 
SALT,
under programs 2013-1-HET-005 (31\% complete, PI Jason Curtis) and 
2013-2-HET-003 (PI Jason Curtis). 
We thank SALT's support astronomers and staff for their assistance, 
especially Petri Vaisanen and Alexei Kniazev.

This publication makes use of data products from the Two Micron All Sky 
Survey, which is a joint project of the University of Massachusetts and 
the Infrared Processing and Analysis Center/California Institute of 
Technology, funded by the National Aeronautics and Space Administration 
and the National Science Foundation.
This research also made use of the WEBDA database, once operated at the Institute for Astronomy of the University of Vienna,
now operated at the Department of Theoretical Physics and Astrophysics of the Masaryk University; 
NASA’s Astrophysics Data System Bibliographic Services;  
the SIMBAD database and the VizieR catalog access tool operated at CDS, Strasbourg, France; 
and the VizieR database \citep{vizier}.
This research made use of Montage, funded by NASA’s Earth
Science Technology Office, Computation Technologies Project,
under Cooperative Agreement Number NCC5-626 between
NASA and the California Institute of Technology. Montage is
maintained by the NASA/IPAC Infrared Science Archive.
This research was made possible through the use of the 
AAVSO Photometric All-Sky Survey (APASS), 
funded by the Robert Martin Ayers Sciences Fund.

{\it Facilities:} \facility{SALT (RSS)}


\bibliographystyle{apj}

\begin{thebibliography}{43}
\expandafter\ifx\csname natexlab\endcsname\relax\def\natexlab#1{#1}\fi

\bibitem[{Allen \& Cox(2000)}]{AllenAstroQuant}
Allen, C., \& Cox, A. 2000, Allen's Astrophysical Quantities (Springer)

\bibitem[{{Barnes}(2003)}]{Barnes2003}
{Barnes}, S.~A. 2003, \apj, 586, 464

\bibitem[{{Becker} {et~al.}(2015){Becker}, {Johnson}, {Vanderburg}, \&
  {Morton}}]{Becker2015}
{Becker}, J.~C., {Johnson}, J.~A., {Vanderburg}, A., \& {Morton}, T.~D. 2015,
  \apjs, 217, 29

\bibitem[{{Bowsher} {et~al.}(2012){Bowsher}, {Ag{\"u}eros}, {Bochanski},
  {Cargile}, {Covey}, {Kraus}, {Law}, \& {Stassun}}]{Bowsher752}
{Bowsher}, E.~C., {Ag{\"u}eros}, M., {Bochanski}, J., {Cargile}, P., {Covey},
  K., {Kraus}, A., {Law}, N., \& {Stassun}, K. 2012, in American Astronomical
  Society Meeting Abstracts, Vol. 219, American Astronomical Society Meeting
  Abstracts \#219, 151.25


\bibitem[Brandt et al.(2001)]{Brandt2001} Brandt, W.~N., Alexander, D.~M., Hornschemeier, A.~E., et al.\ 2001, \aj, 122, 2810


\bibitem[Bressan et al.(2012)]{parsec} Bressan, A., Marigo, P., Girardi, L., et al.\ 2012, \mnras, 427, 127

\bibitem[Buckley et al.(2006)]{salt1} Buckley, D.~A.~H., Swart, G.~P., \& Meiring, J.~G.\ 2006, \procspie, 6267, 62670Z

\bibitem[{{Burgh} {et~al.}(2003){Burgh}, {Nordsieck}, {Kobulnicky}, {Williams},
  {O'Donoghue}, {Smith}, \& {Percival}}]{rss1}
{Burgh}, E.~B., {Nordsieck}, K.~H., {Kobulnicky}, H.~A., {Williams}, T.~B.,
  {O'Donoghue}, D., {Smith}, M.~P., \& {Percival}, J.~W. 2003, in Proc. SPIE, Vol. 4841,
  Instrument Design and Performance for Optical/Infrared Ground-based
  Telescopes, ed. M.~{Iye} \& A.~F.~M. {Moorwood}, 1463--1471

\bibitem[Burnham(1980)]{burnham} Burnham, R.\ 1980, Mercury, 9, 19


\bibitem[{{Chubak} {et~al.}(2012){Chubak}, {Marcy}, {Fischer}, {Howard},
  {Isaacson}, {Johnson}, \& {Wright}}]{carly}
{Chubak}, C., {Marcy}, G., {Fischer}, D.~A., {Howard}, A.~W., {Isaacson}, H.,
  {Johnson}, J.~A., \& {Wright}, J.~T. 2012, arXiv:1207.6212


\bibitem[Crawford et al.(2010)]{pysalt} Crawford, S.~M., Still, M., Schellart, P., et al.\ 2010, \procspie, 7737, 773725

\bibitem[{{Curtis} {et~al.}(2013){Curtis}, {Wolfgang}, {Wright}, {Brewer}, \&
  {Johnson}}]{Curtis2013}
{Curtis}, J.~L., {Wolfgang}, A., {Wright}, J.~T., {Brewer}, J.~M., \&
  {Johnson}, J.~A. 2013, \aj, 145, 134

\bibitem[{{Dias} {et~al.}(2002){Dias}, {Alessi}, {Moitinho}, \&
  {L{\'e}pine}}]{dias2002}
{Dias}, W.~S., {Alessi}, B.~S., {Moitinho}, A., \& {L{\'e}pine}, J.~R.~D. 2002,
  \aap, 389, 871

\bibitem[{{Dotter} {et~al.}(2008){Dotter}, {Chaboyer}, {Jevremovi{\'c}},
  {Kostov}, {Baron}, \& {Ferguson}}]{dartmouth}
{Dotter}, A., {Chaboyer}, B., {Jevremovi{\'c}}, D., {Kostov}, V., {Baron}, E.,
  \& {Ferguson}, J.~W. 2008, \apjs, 178, 89


\bibitem[Frinchaboy et al.(2013)]{OCCAM} Frinchaboy, P.~M., Thompson, B., Jackson, K.~M., et al.\ 2013, \apjl, 777, L1

\bibitem[{{Giardino} {et~al.}(2008){Giardino}, {Pillitteri}, {Favata}, \&
  {Micela}}]{Giardino2008}
{Giardino}, G., {Pillitteri}, I., {Favata}, F., \& {Micela}, G. 2008, \aap,
  490, 113

\bibitem[{{Han} {et~al.}(2014){Han}, {Curtis}, \& {Wright}}]{eunkyuposter}
{Han}, E., {Curtis}, J.~L., \& {Wright}, J. 2014, in American Astronomical
  Society Meeting Abstracts, Vol. 223, American Astronomical Society Meeting
  Abstracts \#223, \#442.06

\bibitem[{{Herschel}(1833)}]{firstr147ref}
{Herschel}, J.~F.~W. 1833, Philosophical Transactions of the Royal Society of
  London, 123, 359

\bibitem[H{\o}g et al.(2000)]{tycho2} H{\o}g, E., Fabricius, C., Makarov, V.~V., et al.\ 2000, \aap, 355, L27

\bibitem[{{Hooper} {et~al.}(2012){Hooper}, {Nordsieck}, {Williams}, {Buckley},
  {SALT Operations Group}, \& {UW-Madison RSS Commissioning Group}}]{rss3}
{Hooper}, E.~J., {Nordsieck}, K., {Williams}, T., {Buckley}, D., {SALT
  Operations Group}, \& {UW-Madison RSS Commissioning Group}. 2012, in American
  Astronomical Society Meeting Abstracts, Vol. 219, American Astronomical
  Society Meeting Abstracts \#219, \#422.10

\bibitem[Howell et al.(2014)]{Howell2014} Howell, S.~B., Sobeck, C., Haas, M., et al.\ 2014, \pasp, 126, 398

\bibitem[{{Johnson} \& {Sandage}(1955)}]{classicM67}
{Johnson}, H.~L., \& {Sandage}, A.~R. 1955, \apj, 121, 616

\bibitem[Kharchenko et al.(2004)]{khar2004} 
Kharchenko, N.~V., Piskunov, A.~E., R{\"o}ser, S., Schilbach, E., 
\& Scholz, R.-D.\ 2004, AN, 325, 740 

\bibitem[{{Kharchenko} {et~al.}(2005){Kharchenko}, {Piskunov}, {R{\"o}ser},
  {Schilbach}, \& {Scholz}}]{khar2005}
{Kharchenko}, N.~V., {Piskunov}, A.~E., {R{\"o}ser}, S., {Schilbach}, E., \&
  {Scholz}, R. 2005, \aap, 438, 1163

\bibitem[{{Kharchenko} {et~al.}(2013){Kharchenko}, {Piskunov}, {Schilbach},
  {R{\"o}ser}, \& {Scholz}}]{khar13}
{Kharchenko}, N.~V., {Piskunov}, A.~E., {Schilbach}, E., {R{\"o}ser}, S., \&
  {Scholz}, R.-D. 2013, \aap, 558, A53

\bibitem[{{Kobulnicky} {et~al.}(2003){Kobulnicky}, {Nordsieck}, {Burgh},
  {Smith}, {Percival}, {Williams}, \& {O'Donoghue}}]{rss2}
{Kobulnicky}, H.~A., {Nordsieck}, K.~H., {Burgh}, E.~B., {Smith}, M.~P.,
  {Percival}, J.~W., {Williams}, T.~B., \& {O'Donoghue}, D. 2003, in Society of
  Photo-Optical Instrumentation Engineers (SPIE) Conference Series, Vol. 4841,
  Instrument Design and Performance for Optical/Infrared Ground-based
  Telescopes, ed. M.~{Iye} \& A.~F.~M. {Moorwood}, 1634--1644

\bibitem[{{Landsman}(1993)}]{idlastro}
{Landsman}, W.~B. 1993, in Astronomical Society of the Pacific Conference
  Series, Vol.~52, Astronomical Data Analysis Software and Systems II, ed.
  R.~J. {Hanisch}, R.~J.~V. {Brissenden}, \& J.~{Barnes}, 246

\bibitem[{{Lod{\'e}n}(1980)}]{loden-mem}
{Lod{\'e}n}, L.~O. 1980, \aaps, 41, 173

\bibitem[{{Mamajek} \& {Hillenbrand}(2008)}]{mamajek2008}
{Mamajek}, E.~E., \& {Hillenbrand}, L.~A. 2008, \apj, 687, 1264

\bibitem[{{Mermilliod} \& {Paunzen}(2003)}]{webda}
{Mermilliod}, J., \& {Paunzen}, E. 2003, \aap, 410, 511

\bibitem[{{Ochsenbein} {et~al.}(2000){Ochsenbein}, {Bauer}, \&
  {Marcout}}]{vizier}
{Ochsenbein}, F., {Bauer}, P., \& {Marcout}, J. 2000, \aaps, 143, 23

\bibitem[O'Donoghue et al.(2006)]{salt2} O'Donoghue, D., Buckley, D.~A.~H., Balona, L.~A., et al.\ 2006, \mnras, 372, 151

\bibitem[{{Prusti}(2012)}]{gaia}
{Prusti}, T. 2012, Astronomische Nachrichten, 333, 453

\bibitem[Ramsey et al.(1998)]{HET} Ramsey, L.~W., Adams, M.~T., Barnes, T.~G., et al.\ 1998, \procspie, 3352, 34


\bibitem[{{Robin} {et~al.}(2003){Robin}, {Reyl{\'e}}, {Derri{\`e}re}, \&
  {Picaud}}]{galsim}
{Robin}, A.~C., {Reyl{\'e}}, C., {Derri{\`e}re}, S., \& {Picaud}, S. 2003,
  \aap, 409, 523

\bibitem[{{Roeser} {et~al.}(2010){Roeser}, {Demleitner}, \&
  {Schilbach}}]{ppmxl}
{Roeser}, S., {Demleitner}, M., \& {Schilbach}, E. 2010, \aj, 139, 2440

\bibitem[R{\"o}ser \& Bastian(1994)]{ngc2451} R{\"o}ser, S., \& Bastian, U.\ 1994, \aap, 285

\bibitem[{{Saar} {et~al.}(2014){Saar}, {Curtis}, \& {Wright}}]{SaarChandra}
{Saar}, S.~H., {Curtis}, J.~L., \& {Wright}, J. 2014, in American Astronomical
  Society Meeting Abstracts, Vol. 224, American Astronomical Society Meeting
  Abstracts \#224, 322.03

\bibitem[{{Soderblom}(2010)}]{soderblom2010}
{Soderblom}, D.~R. 2010, \araa, 48, 581

\bibitem[{{Sulentic} {et~al.}(1973){Sulentic}, {Tifft}, \&
  {Dreyer}}]{revisedNewCatalog}
{Sulentic}, J.~W., {Tifft}, W.~G., \& {Dreyer}, J.~L.~E. 1973, {The Revised New
  Catalogue of Nonstellar Astronomical Objects}

\bibitem[{{Tonry} \& {Davis}(1979)}]{rvs}
{Tonry}, J., \& {Davis}, M. 1979, \aj, 84, 1511

\bibitem[Twarog et al.(2015)]{Twarog752} Twarog, B.~A., Anthony-Twarog, B.~J., Deliyannis, C.~P., \& Thomas, D.~T.\ 2015, \aj, 150, 134


\bibitem[{{Wallace} {et~al.}(2011){Wallace}, {Hinkle}, {Livingston}, \&
  {Davis}}]{wallace2011}
{Wallace}, L., {Hinkle}, K.~H., {Livingston}, W.~C., \& {Davis}, S.~P. 2011,
  \apjs, 195, 6

\bibitem[{{Zacharias} {et~al.}(2012){Zacharias}, {Finch}, {Girard}, {Henden},
  {Bartlett}, {Monet}, \& {Zacharias}}]{ucac4}
{Zacharias}, N., {Finch}, C.~T., {Girard}, T.~M., {Henden}, A., {Bartlett},
  J.~L., {Monet}, D.~G., \& {Zacharias}, M.~I. 2012, VizieR Online Data
  Catalog, 1322, 0

\end{thebibliography}

\end{document}